%% file: jhep_scan_v4.6.tex
\documentclass[a4paper,11pt]{article}

\usepackage{jheppub} 

\usepackage[T1]{fontenc} 

\usepackage{multirow}
 
\usepackage{rotating}
\usepackage{longtable}
\usepackage{dcolumn}  

\usepackage{booktabs}

\renewcommand{\arraystretch}{1.1}

\newcommand{\mm}{M_{\rm recoil}}

\newcommand{\ev}{\mathrm{eV}}

\newcommand{\mev}{\mathrm{MeV}}

\newcommand{\mevm}{\mathrm{MeV}/c^2}
\newcommand{\gev}{\mathrm{GeV}}

\newcommand{\gevm}{\mathrm{GeV}/c^2}

\newcommand{\ee}{e^+e^-}
\newcommand{\uu}{\mu^+\mu^-}
\newcommand{\pp}{\pi^+\pi^-}

\newcommand{\U}{\Upsilon}
\newcommand{\Uf}{\Upsilon(10860)}
\newcommand{\Us}{\Upsilon(11020)}
\newcommand{\Uo}{\Upsilon(1S)}

\newcommand{\Ut}{\Upsilon(2S)}
\newcommand{\Uth}{\Upsilon(3S)}
\newcommand{\Un}{\Upsilon(nS)}
\newcommand{\muu}{M(\mu^+\mu^-)}
\newcommand{\mmpp}{M_{\rm recoil}(\pi^+\pi^-)}

\newcommand{\mpp}{M(\pi^+\pi^-)}
\newcommand{\mppsq}{M^2(\pi^+\pi^-)}

\newcommand{\hbn}{h_b(nP)}

\newcommand{\ks}{K^0_S}

\newcommand{\fb}{\mathrm{fb}^{-1}}
\newcommand{\br}{\mathcal{B}}
\newcommand{\lik}{\mathcal{L}}

\newcommand{\ecm}{E_{\rm c.m.}}

\newcommand{\zbo}{Z_b(10610)}
\newcommand{\zbt}{Z_b(10650)}
\newcommand{\gee}{\Gamma_{\rm ee}}

\title{\boldmath Observation of a new structure near $10.75\,\gev$
  in the energy dependence of the $e^+e^-\to\Upsilon(nS)\pi^+\pi^-$
  ($n=1,2,3$) cross sections}

\input{pub543}

\abstract{ We report a new measurement of the
  $e^+e^-\to\Upsilon(nS)\pi^+\pi^-$ ($n=1,2,3$) cross sections at
  energies from $10.52$ to $11.02\,\gev$ using data collected with the
  Belle detector at the KEKB asymmetric-energy $\ee$ collider. 
  We observe a new structure in the energy dependence of the cross 
  sections; if described by a Breit-Wigner function its mass and width
  are found to be $M=(10752.7\pm5.9\,^{+0.7}_{-1.1})\,\mevm$ and
  $\Gamma=(35.5^{+17.6}_{-11.3}\,^{+3.9}_{-3.3})\,\mev$, where the
  first error is statistical and the second is systematic. The global
  significance of the new structure including systematic
  uncertainty is 5.2 standard deviations.
  We also find evidence for the $\ee\to\Uo\pp$ process at the energy
  $10.52\,\gev$, which is below the $B\bar{B}$ threshold.  }

\keywords{e+e- Experiments, Quarkonium, Spectroscopy}

\begin{document}
\maketitle
\flushbottom

\section{Introduction}

The observed 
vector bottomonium states above the $B\bar{B}$ threshold, $\U(4S)$,
$\Uf$, and $\Us$, have properties that are unexpected for pure
$b\bar{b}$ bound states~\cite{Bondar:2016hva}. Their transitions to
lower bottomonia with the emission of light hadrons have much higher
rates compared to expectations for ordinary bottomonium, and some of
these transitions strongly violate Heavy Quark Spin Symmetry. A
possible explanation of these unusual properties is a contribution of
hadron loops or, equivalently, the presence of a
$B_{(s)}^{(*)}\bar{B}_{(s)}^{(*)}$ admixture in the bottomonium wave
function~\cite{Meng:2007tk,Simonov:2008ci,Voloshin:2012dk}. In this
approach, the $\Uf$ and $\Us$ are the $\U(5S)$ and $\U(6S)$ states
``dressed'' by hadrons.

In the region of the $\U(4S,5S,6S)$ states, quark models also predict
the $\U(3D,4D)$ levels~\cite{Ebert:2011jc,Godfrey:2015dia}. The
electron widths of the $D$-wave states arise from mixing with the
$S$-wave states, which is expected to be quite small for bottomonia
below the $B\bar{B}$ threshold~\cite{Moxhay:1983vu}. However, it can
be significantly enhanced for the states above open-flavor thresholds
because of $B$-meson loops~\cite{Badalian:2009bu}. Thus, the
``dressed'' $\U(3D,4D)$ states might be formed abundantly in $\ee$
annihilations.

The unexpected properties of the $\Uf$ and $\Us$ could also be due to
the presence of other exotic states, e.g., compact
tetraquarks~\cite{Ali:2009es} or
hadrobottomonia~\cite{Dubynskiy:2008mq}. To understand the nature of
the already known $\Upsilon$ states above the $B\bar{B}$ threshold and
to search for additional states expected in this energy region, it is of
interest to study the energy dependence of the $\ee\to(b\bar{b}...)$
cross sections, where $(b\bar{b}...)$ denotes exclusive final states
containing the $b$ and $\bar{b}$ quarks, both with open and hidden
flavor.

Recently, the Belle experiment measured the energy dependence of the
cross sections $\ee\to\Un\pp$ ($n=1,2,3$)~\cite{Santel:2015qga},
$\ee\to\hbn\pp$ ($n=1,2$)~\cite{Abdesselam:2015zza},
$\ee\to\chi_{bJ}(1P)\,\pp\pi^0$~\cite{Yin:2018ojs}, and
$\ee\to{B}_s^{(*)}\bar{B}_s^{(*)}$~\cite{Abdesselam:2016tbc} in the
energy region from 10.63 to $11.02\,\gev$. The shapes of the $\Un\pp$
and $\hbn\pp$ cross sections show prominent $\Uf$ and $\Us$ signals
with no significant non-resonant contributions. Production of the
$\hbn\pp$ is found to proceed entirely via the exotic charged
bottomonium-like states $\zbo$ and $\zbt$~\cite{Belle:2011aa}. The
$B_s^*\bar{B}_s^*$ cross section shows a prominent $\Uf$ signal, while
the $B_s\bar{B}_s$ and $B_s\bar{B}_s^*$ cross sections are relatively
small and do not show any significant structures. The uncertainties in
the $\chi_{bJ}(1P)\,\pp\pi^0$ cross section at various energies are
too large to draw conclusions about its shape. No evidence is found
for new structures in any of these cross sections except possibly near
$10.77\,\gev$ in the $\Un\pp$ final states.
It is of interest to study more channels and to improve the accuracy
of the previously measured cross sections.

In this paper, we report an updated measurement of the
$\sigma[\ee\to\Un\pp]$ ($n=1,2,3$) energy dependence. We improve the
accuracy by reconstructing $\Un\to\ee$ in addition to $\uu$, and by
extracting signal yields via fits to the $\pp$ recoil mass
distributions, instead of counting events with inverse efficiency
weights in the signal and sideband regions, as was done
previously~\cite{Santel:2015qga}. We also use the
initial-state-radiation (ISR) process in the high statistics $\Uf$
on-resonance data to obtain additional information about the cross
section shapes.
As a result of these improvements, we observe a new structure in the
energy dependence of the $\ee\to\Un\pp$ ($n=1,2,3$) cross sections
with a mass near $10.75\,\gev$.
We also find evidence for the $\ee\to\Uo\pp$ process below the
$B\bar{B}$ threshold at $10.52\,\gev$. This implies that the
$\ee\to\Un\pp$ cross sections have non-resonant contributions.

The paper is organized as follows. In Section~\ref{sec:data_sets} we
present the data sets used in the analysis, and briefly describe the Belle
detector and Monte-Carlo (MC) simulation. In
Section~\ref{sec:selection} we list selection requirements for signal
events. Section~\ref{sec:eeuu_calib} is devoted to the center-of-mass
(c.m.) energy 
calibration using the $\ee\to\uu$ process. In Section~\ref{sec:shape}
we present an analytical calculation of the $\ee\to\Un\pp$ signal
shapes in the $\mmpp$ distributions and calibration of momentum
resolution using the $\U(2S,3S)\to\U(1S,2S)\pp$ transitions in the
$\Ut$ and $\Uth$ data. Fits to the $\mmpp$ distributions are described
in Section~\ref{sec:mmpp_fits}, while the results of the measurement of
the Born cross sections and c.m.\ energy calibration using combined
$\ee\to\uu$ and $\ee\to\Un\pp$ processes are presented in
Section~\ref{sec:born}. The fit to the cross section energy
dependence, determination of the parameters of the intermediate states, 
and of the significance of the new structure are presented
in Section~\ref{sec:fit}. In Section~\ref{sec:concl} we give
conclusions and mention possible interpretations of the new structure.

\section{Data sets, Belle detector and simulation}
\label{sec:data_sets}

The analysis is based on data collected by the Belle detector at
the KEKB asymmetric-energy $\ee$ collider~\cite{BELLE_DETECTOR,KEKB}.
We use energy scan data with approximately $1\,\fb$ per point
collected in the energy range from $10.63\,\gev$ to
$11.02\,\gev$. These data were collected during two running periods,
one with six energy points in 2007 and the other with sixteen energy
points in 2010. We also use the $\Uf$ on-resonance data sample, with a
total luminosity of $121\,\fb$. These data were collected in five
running periods with slightly different c.m.\ energies between
$10.864\,\gev$ and $10.868\,\gev$. Finally, we use $60\,\fb$ of the
continuum data sample collected at $10.52\,\gev$. Thus, in total,
there are 28 energy points where we calibrate c.m.\ energies and
measure cross sections. Energies and luminosities of various data
samples are presented in Table~\ref{tab:ecm_lu_bo}.

The Belle detector is a large-solid-angle magnetic spectrometer that
consists of a silicon vertex detector (SVD), a 50-layer central drift
chamber (CDC), an array of aerogel threshold Cherenkov counters (ACC),
a barrel-like arrangement of time-of-flight scintillation counters
(TOF), and an electromagnetic calorimeter (ECL) comprised of CsI(Tl)
crystals located inside a superconducting solenoid that provides a
1.5~T magnetic field. An iron flux return located outside the coil is
instrumented to detect $K_L^0$ mesons and to identify muons (KLM). A
more detailed description of the detector can be found in
Ref.~\cite{BELLE_DETECTOR}.

The integrated luminosity is measured with barrel Bhabha
events. The systematic error in the luminosity measurement is about
1.4\% and is dominated by the theoretical uncertainty in the Bhabha
cross section; the statistical error is usually small compared to
the systematic error.

The detector response is simulated using GEANT~\cite{GEANT}. The MC
simulation includes run-dependent variations in the detector performance
and background conditions. 
The $\ee\to\Un\pp$ ($n=1,2,3$) events including ISR are generated with
EvtGen~\cite{Lange:2001uf}. We use matrix elements measured in the
$\Uf$ on-resonance data~\cite{Garmash:2014dhx}. 
For the energy points outside the $\Uf$ peak, we use uniform Dalitz
plot (DP) distributions to assess systematic uncertainty.
The process $\ee\to\uu$ that is used in the c.m.\ energy calibration
is simulated with Phokhara~\cite{PHOKHARA}.
For the $\U(2S,3S)\to\U(2S,1S)\pp$ transitions we use matrix elements
measured by the CLEO experiment~\cite{CroninHennessy:2007zz}.
Background from QED production of four-track final states is
simulated using a specially developed extension of the BDK
generator~\cite{Krachkov:2019kty}.
Final-state radiation (FSR) is simulated with
PHOTOS~\cite{PHOTOS}.

\section{Event selection}
\label{sec:selection}

We select events of the type $\ee\to\Un\pp$ ($n=1,2,3$) with
$\Un\to\uu$ or $\ee$.
A preselected sample containing $\uu$ or $\ee$ pairs with invariant
masses greater than $8\,\gevm$ is used.
Muons are identified by their range and transverse scattering in the
KLM~\cite{Abashian:2002bd}.
Electrons are identified by the presence of an ECL cluster matching a
track in position and energy and having a transverse energy profile
consistent with an electromagnetic shower; the ionization loss
measurement in the CDC and the response of the ACC are also
used~\cite{Nakano:2002jw}.
We require the presence of two additional charged tracks that are
identified as pions. Identification is based on the information from
the TOF and ACC, combined with the ionization loss measurement in the
CDC. We also apply an electron veto. The total efficiency of the
identification requirements is at the level of 99\% per pion.
All tracks are required to originate from the interaction point (IP)
region; this requirement helps eliminate poorly reconstructed
tracks.
Multiple candidates occur in about 0.5\% of the events. We select the
best candidate based on the smallest distance to IP in the plane
transverse to the beam direction.

To suppress background from converted photons, we require
$\cos\theta_{\pp}<0.95$, where $\theta_{\pp}$ is the pion opening
angle in the laboratory frame.
In the $\ee\pp$ final state we apply additional requirements,
$M_{\mathrm{recoil}}(\ee)>350\,\mevm$ and $\cos\theta_{e^-}<0.82$,
where $\theta_{e^-}$ is the angle between the $e^-$ momentum in the
c.m.\ frame and the electron beam.
The $\ee$ recoil mass is defined as
\begin{equation}
  \mm(\ee)=\sqrt{(\ecm-E_{\ee})^2 - p_{\ee}^{\phantom{i}2}},
\end{equation}
where $\ecm$ is the c.m.\ energy, $E_{\ee}$ and $p_{\ee}$ are the
$\ee$ energy and momentum measured in the c.m.\ frame.

\begin{figure}[htbp]
\centering
\includegraphics[width=0.49\textwidth]{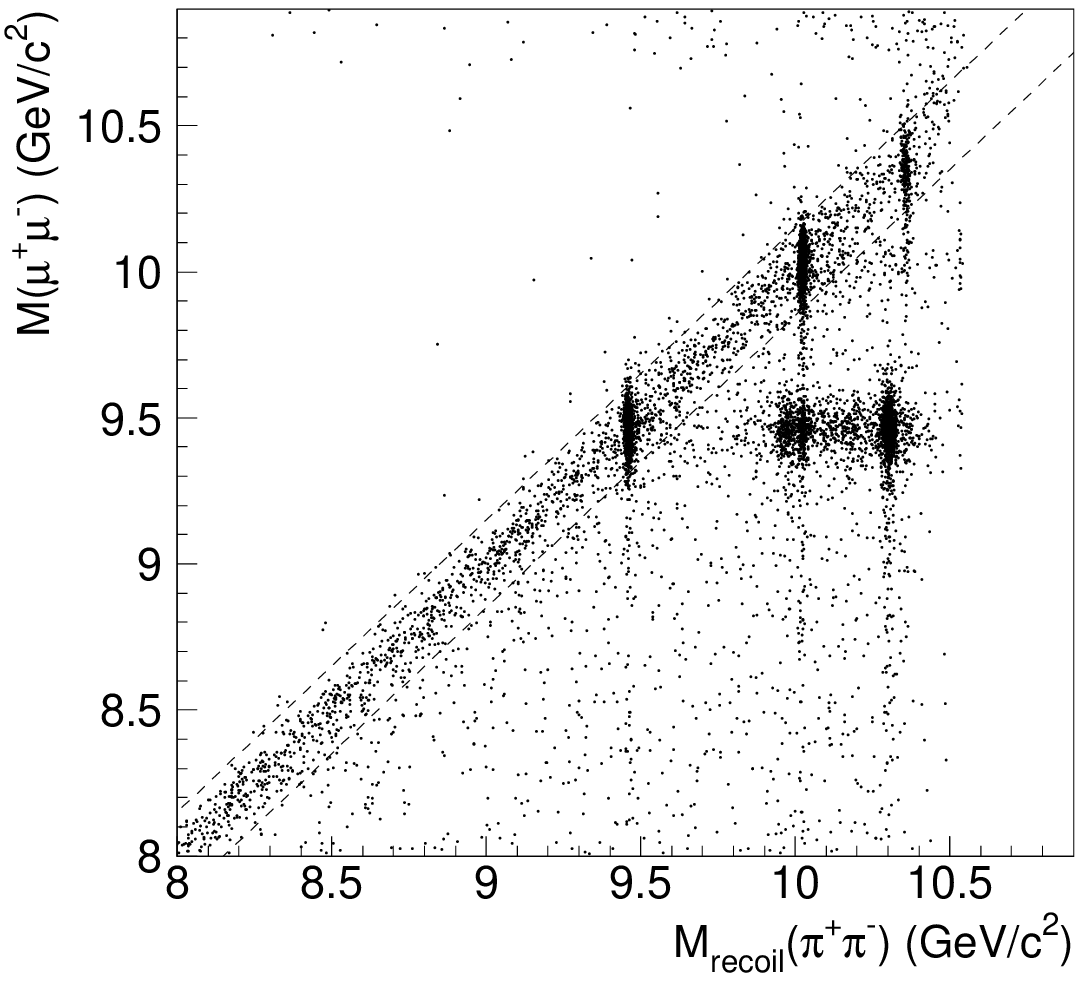}\hfill
\includegraphics[width=0.49\textwidth]{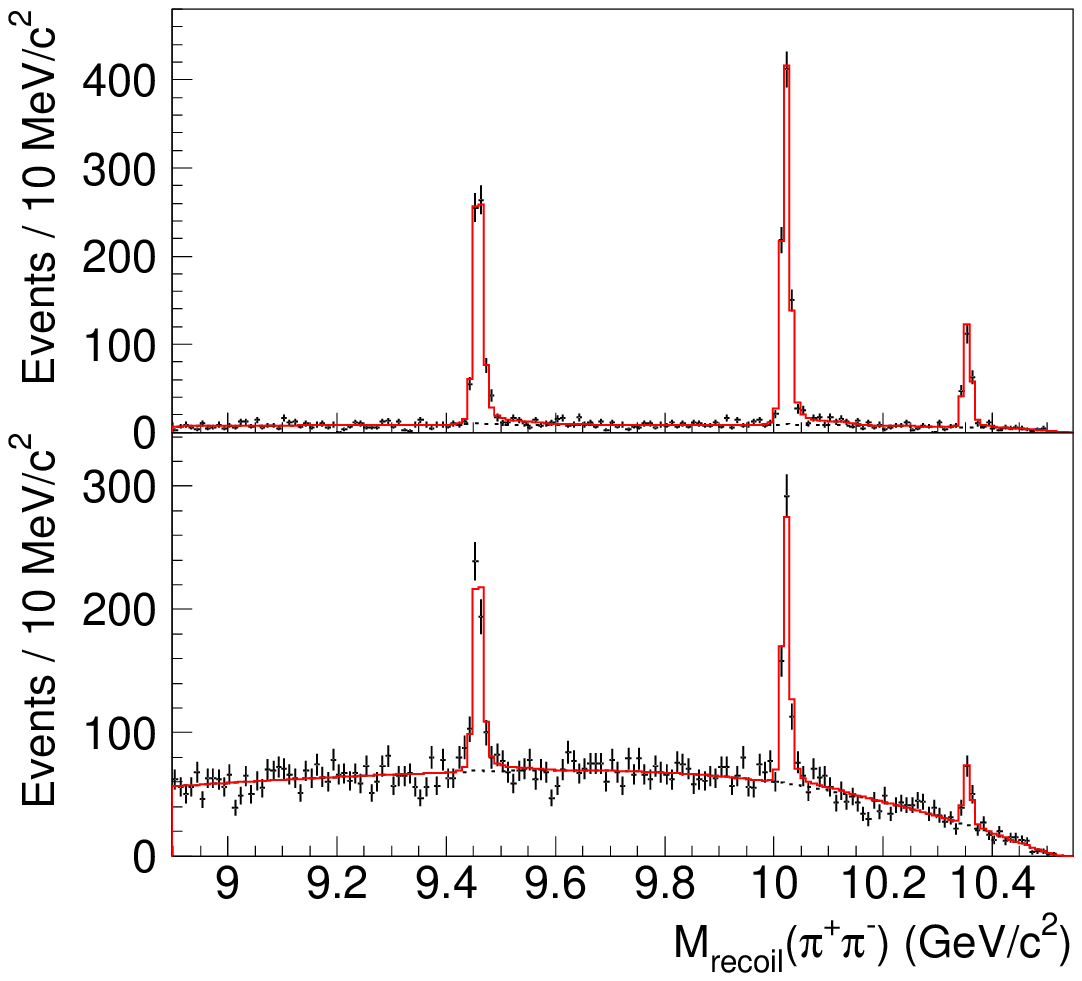}
\caption{ Left: $\muu$ vs.\  $\mmpp$ distribution in a 
 high-statistics data sample (No.\ 9 in Table~\ref{tab:ecm_lu_bo}).
  Dashed lines show a diagonal band selected by the requirements of
  Eq.~(\ref{eq:ene_balance}).
  Right: $\mmpp$ distributions for the selected $\uu\pp$ (top) and
  $\ee\pp$ (bottom) candidates. Results of the fit (red solid
  histogram) and background component of the fit function (black dotted
  histogram) are also shown. }
\label{muu_vs_mmpp_exp69_051218}
\end{figure}

Figure~\ref{muu_vs_mmpp_exp69_051218} (left) shows the $\muu$
vs.\ $\mmpp$ distribution for a high-statistics data sample (No.\ 9 in
Table~\ref{tab:ecm_lu_bo}).
The clusters along the diagonal are due to $\ee\to\Un\pp$ ($n=1,2,3$).
Below the diagonal there are events in which some final-state particles
are not reconstructed.
The fully reconstructed events are selected with a requirement:
\begin{equation}
  -150<\mmpp-M(\ell^+\ell^-)<150\,\mev,
  \label{eq:ene_balance}
\end{equation}
where $\ell=\mu$ or $e$.
Since $\mmpp\approx\ecm-E_{\pp}$ and $\muu\approx E_{\uu}$,
Eq.~(\ref{eq:ene_balance}) corresponds to an energy
balance requirement.
The $\mmpp$ distributions are presented in
Fig.~\ref{muu_vs_mmpp_exp69_051218} (right) for both $\uu\pp$ and
$\ee\pp$ events.

\section{\boldmath Calibration of the center-of-mass energies with
  $\ee\to\uu$} 
\label{sec:eeuu_calib}

In this section we describe the c.m.\ energy calibration with
$\ee\to\uu$. Subsequently, the $\ecm$ calibration is improved using
the $\ee\to\Un\pp$ ($n=1,2,3$) processes, as described in
Section~\ref{sec:mmpp_fits}. 
For the $\Uf$ on-resonance data we use only $\ee\to\Un\pp$.

\begin{figure}[htbp]
\centering
\includegraphics[width=0.49\textwidth]{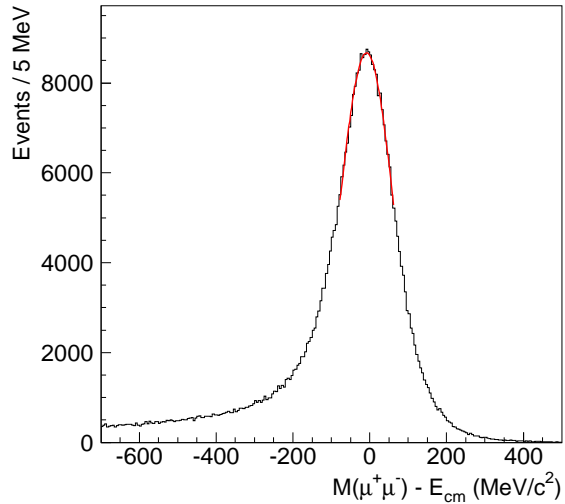}
\caption{ $M(\uu)-\ecm$ distribution for selected $\ee\to\uu$
  candidates in the scan data sample. Red curve is the fit result. }
\label{fit_mupair_e73_16}
\end{figure}

We select $\ee\to\uu$ events with the same requirements as described
in Section~\ref{sec:selection} except requiring the presence of the $\pp$ 
pair. We fit the $M(\uu)-\ecm$ distributions for all data
samples to a Gaussian in the range $\pm70\,\mevm$ from the peak
position, which corresponds to about $\pm1\,\sigma$. Here and
throughout the paper, we use a binned maximum-likelihood fit unless
stated otherwise. An example of a fitted $M(\uu)$ distribution in the
scan data sample is shown in Fig.~\ref{fit_mupair_e73_16}.
The statistical uncertainty in the peak position is typically
$0.3\,\mevm$.

There is a difference between the $\muu$ peak position and $\ecm$ due
to radiative effects. This difference is determined to be
$-8.3\pm0.6\,\mev$ using the $\ee\to\Un\pp$ processes in the $\Uf$
on-resonance data as described in Section~\ref{sec:mmpp_fits}. The
error assigned is based on the scatter of the measurements in
different on-resonance data samples; it corresponds to the uncertainty
due to long-term stability of the detector performance.
The dependence of the shift on $\ecm$ is determined from MC
simulation. The shift increases in absolute value by $0.3\,\mev$ over
the energy scan range $10.52-11.02\,\gev$.

To estimate the systematic uncertainty in $\ecm$, we vary the fit
interval from $\pm50\,\mevm$ to $\pm80\,\mevm$ and take the largest
deviation as an error; it is in the range $0.1-0.7\,\mev$. We
investigate the stability of the $\ks\to\pp$ invariant mass over the
data-taking period and assign a $1\,\mev$ uncertainty due to a
possible variation of the Belle solenoid magnetic field during energy
scan. The overall uncertainty is determined as a quadrature sum of all
contributions; the values are in the range $1.2-1.4\,\mev$.

\section{\boldmath Signal density functions in the $\mmpp$ distributions}
\label{sec:shape}

The c.m.\ energy calibration using the $\ee\to\Un\pp$ ($n=1,2,3$)
processes and the measurement of the corresponding yields is done by
fitting the $\pp$ recoil mass distributions. In this Section we
describe the determination of signal density functions in the $\mmpp$
distributions, which is a key tool of this analysis.

The density functions for the $\Un\pp$ signals in the $\mmpp$
distributions are calculated as a sequence of convolutions that take
into account the $\pp$ momentum resolution, ISR, and the beam
energy spread.

The momentum resolution function is a sum of a symmetric part and a tail on
the high $\mmpp$ side due to FSR, decays in flight, and secondary
interactions. The symmetric part is described by a sum of five
Gaussians. Such a parameterization has enough flexibility to describe
non-Gaussian tails and is fast to compute, which is crucial for
performing convolutions.
The sum of the first three Gaussians contains about 96\% of the
signal, its standard deviation is between $2\,\mevm$ and $4\,\mevm$
for various c.m.\ energies and channels.
The FSR contribution is modeled with a photon energy threshold of
$0.1\,\mev$.

\begin{figure}[htbp]
\centering
\includegraphics[width=0.49\textwidth]{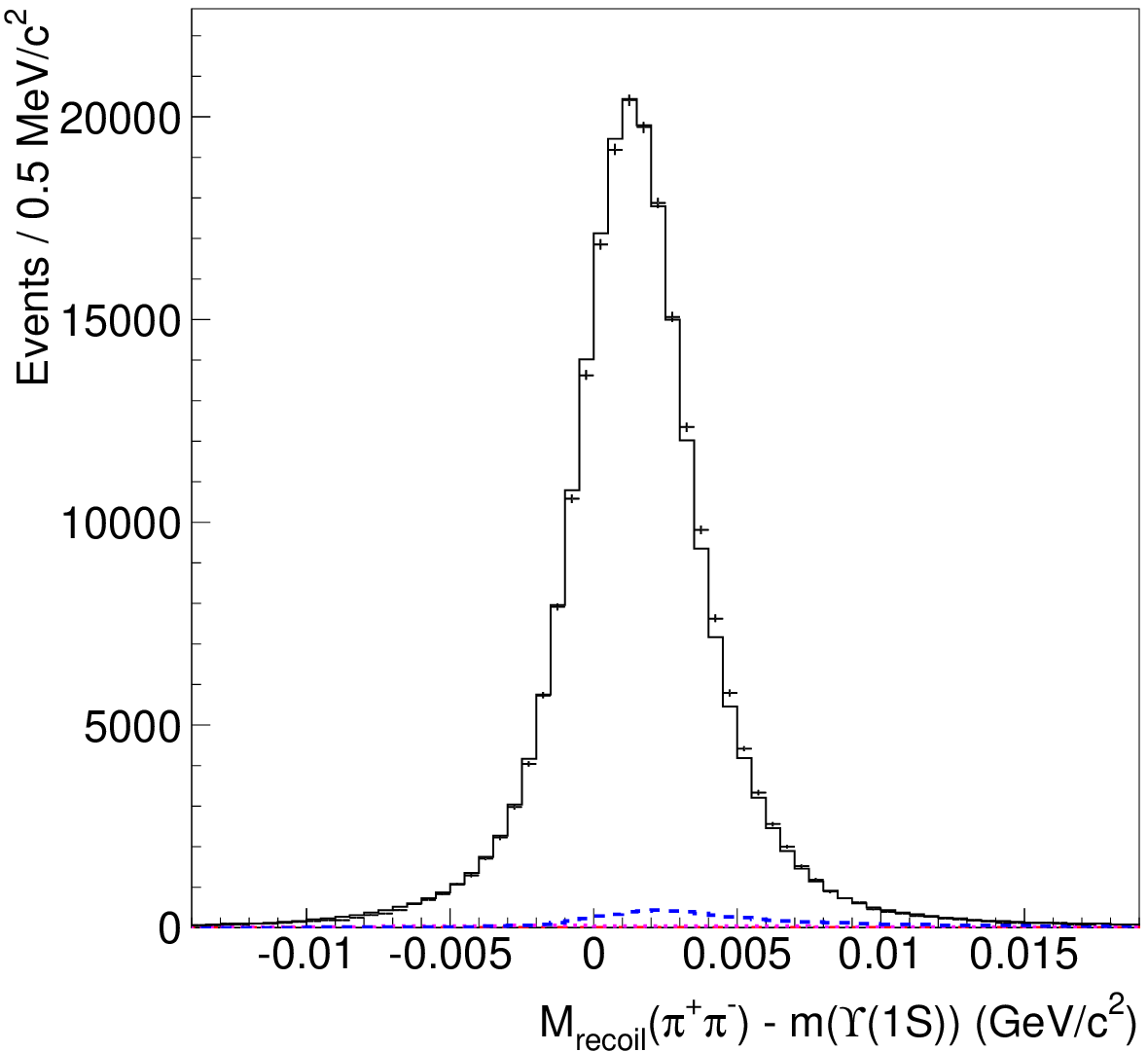}\hfill
\includegraphics[width=0.49\textwidth]{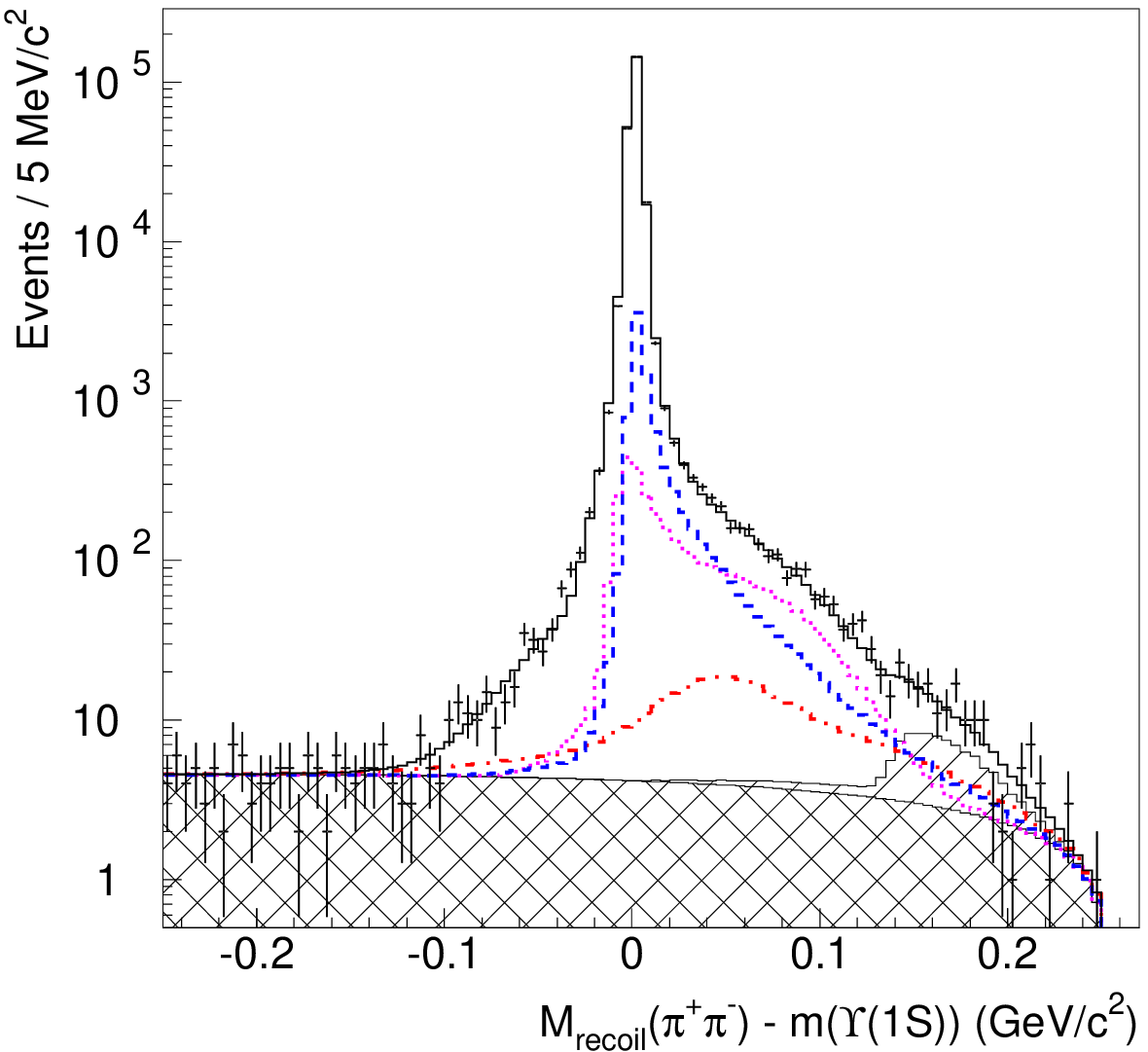}
\caption{ The $\mmpp-m(\Uo)$ distribution for the data collected
  at the $\Ut$ resonance (points with error bars) and fit results
  (black solid histogram). Also shown are combinatorial background
  (cross-hatched histogram), background from $\eta$ decays (hatched
  histogram) and various contributions to the signal function: FSR
  with $\mathrm{E}_{\gamma}>0.1\,\mev$ (blue dashed), decays in flight
  (magenta dotted) and secondary interactions (red dash-dotted). Both
  panels show the same distribution, but differ in bin sizes, linear
  or logarithmic scales of the vertical axes and ranges of the
  horizontal axes. }
\label{y2s_data_liny_301018}
\end{figure}

To calibrate the momentum resolution and to verify the simulation of
its tails we use the high-statistics $\Ut\to\Uo\pp$ signal in the
$25\,\fb$ data sample collected by Belle at the $\Ut$ peak.
We use the same selection requirements as described in
Section~\ref{sec:selection}, with the upper boundary in
Eq.~(\ref{eq:ene_balance}) released to $250\,\mev$. The $\mmpp-m(\Uo)$
distribution is presented in Fig.~\ref{y2s_data_liny_301018}. Due to
the small total width of the $\Ut$ state, the contributions from the
beam energy spread and ISR are negligibly small. Therefore we describe
the signal by the momentum resolution function; its floated parameters
are the normalization, the overall shift in $\mmpp$, and a scale
factor $f$ for the width of the symmetric component (we multiply the
$\sigma$ parameters of all Gaussians by $f$). The non-peaking
background is parameterized with the threshold function
\begin{equation}
  A\,(x-x_0)^p\,e^{-c(x-x_0)},
  \label{eq:argus}
\end{equation}
where $x=\mmpp$; $A$, $x_0$, $p$, and $c$ are parameters that
are floated. We 
also consider the peaking backgrounds due to the $\Ut\to\Uo\eta$
transitions with $\eta\to\pp\pi^0$ and $\pp\gamma$. The shapes of
these peaking backgrounds are determined from MC simulations and their
normalizations are fixed relative to the signal. Fit results are
presented in Fig.~\ref{y2s_data_liny_301018}, which shows that the fit
function describes the data well.
The value of $f$ is found to be
\begin{equation}
  f = 1.160\pm0.003.
  \label{eq:mom_fufa}
\end{equation}
We have used $\Uth\to\U(1S,2S)\pp$ transitions in the $3\,\fb$ data
sample collected at the $\Uth$, and find no indication of a dependence
of $f$ on $\pp$ energy.
The shifts in $\mmpp$ are sensitive to the mass differences
$m(\Ut)-m(\Uo)$, $m(\Uth)-m(\Uo)$, and $m(\Uth)-m(\Ut)$. For all
transitions, we find that the shifts are consistent with zero; thus,
no corrections to the recoil mass scale are applied.

For the ISR probability, we use a calculation to second order in
$\alpha$ by Kuraev and Fadin~\cite{Kuraev:1985hb}. We multiply it by
the relative change of the cross section with $\Delta\ecm$, the shift
in c.m.\ energy due to the emission of a photon, and by the relative
change in the reconstruction efficiency. The efficiency slowly
decreases with $\Delta\ecm$ due to the $\pp$ becoming softer. To
convert from $\Delta\ecm$ to $\Delta\mmpp$ we use the approximate
relation $\Delta\ecm=-\Delta\mmpp$. This approximation
does not produce any visible effects in our analysis.

The contribution of the c.m.\ energy spread is modeled by a Gaussian
multiplied by the cross section energy dependence. The $\sigma$
parameter of the Gaussian is found from a fit to the $\ee\to\Un\pp$
($n=1,2,3$) signals in the combined on-resonance data sample; this fit
is described below. The measured value is
\begin{equation}
  \sigma = 5.36\pm0.13\,\mev,
  \label{eq:spread}
\end{equation}
at the $\Uf$ on-resonance energy of $\ecm=10.866\,\gev$. To find the
spread at other c.m.\ energies, we assume that it is proportional to $\ecm$.
We note that if the cross section changes rapidly with $\ecm$, then, 
because of the energy spread, the average energy of the produced
$\Un\pp$ combination is different from the average energy of the
colliding beams. This can result in a shift of the visible peak
position that is as large as a few $\mevm$.

After the momentum resolution, ISR and energy spread convolutions are
performed, we multiply the resulting signal density function by the
$\Delta\mmpp$ efficiency dependence of the energy balance requirement,
which is a step function smeared by the $M(\uu)$ resolution,
which is typically $\sigma=60\,\mevm$. The smearing is described with
the error function.

The integrals of the momentum resolution and energy spread functions
are normalized to unity; therefore the integral of the signal density
function corresponds to that of the ISR function and the measured signal
yield already includes the ISR correction,
$1+\delta_{\mathrm{ISR}}$. Such a normalization of the signal function
was used in Ref.~\cite{Abdesselam:2015zza}.

The calculation of the signal density function is performed inside the
fit function that allows to float various parameters that influence
the signal, such as the peak position, the c.m.\ energy spread, etc.
The energy dependence of the cross section is required for the
calculation of the signal density function; therefore the analysis is
performed iteratively: we compute the signal density functions using
results of the fit to the cross section energy dependence from the
previous iteration.

\section{\boldmath Fits to $\mmpp$ distributions}
\label{sec:mmpp_fits}

To determine the $\ee\to\Un\pp$ ($n=1,2,3$) signal yields and to
calibrate the c.m.\ energies, we fit the $\mmpp$ distributions in
different data samples. The $\uu\pp$ and $\ee\pp$ candidates are
fitted simultaneously, while the fits in different data samples are
independent.
The fit function is the sum of the signal components, plus non-peaking
and peaking backgrounds.

Signal density functions are calculated as described in
Section~\ref{sec:shape}. The yields in the $\uu\pp$ final state are
floated, while the ratio of the yields in the $\ee\pp$ and $\uu\pp$ is
fixed from MC simulation for all data samples except for the $\Uf$
on-resonance data. This latter sample is used to tune the MC
simulation.
The masses of the $\Uo$, $\Ut$ and $\Uth$ that enter the calculation
of the signal density functions are floated with the constraints of
their world average values~\cite{Patrignani:2016xqp}. We also
introduce a common shift for all peaks, that represents a possible
c.m.\ energy miscalibration.
This common shift is floated without constraints for the on-resonance
data and with the constraints from the $\ee\to\uu$ calibration for the
scan and continuum data.

The $\uu\pp$ final-state background originates from the QED production
of four tracks, e.g. two-photon $\ee$ annihilations in which one
virtual photon produces a high-mass $\uu$ pair, and the other produces
a $\pp$, $\uu$ or $\ee$ pair, with the latter two misidentified as
$\pp$. There is a small peaking component in this type of processes
due to the high-mass lepton pair being produced via the intermediate
$\Un$ ($n=1,2,3$). In case of the $\ee\pp$ final state there are also
photon exchange diagrams and, thus, the background is higher.

The non-peaking component is described by an empirical function 
\begin{equation}
B(x)=A\,(x-x_0)^p\,P_3(x),
\label{eq:bg}
\end{equation}
where $A$, $x_0$ and $p$ are parameters, $P_3(x)$ is a 3rd-order
Chebyshev polynomial with a constant term set to unity. For
high-statistics data samples all background parameters are
floated. For low-statistics samples, only the normalization $A$ is
floated, while the other parameters are fixed using the results of the
fit to the combined on-resonance data sample; the threshold $x_0$ is
recalculated for each of these data samples based on the energy of
each sample.
The peaking background is determined from MC
simulation~\cite{Krachkov:2019kty} separately for each data
sample. Its influence on the measured cross sections is found to be
small.

\begin{figure}[htbp]
\centering
\includegraphics[width=0.49\textwidth]{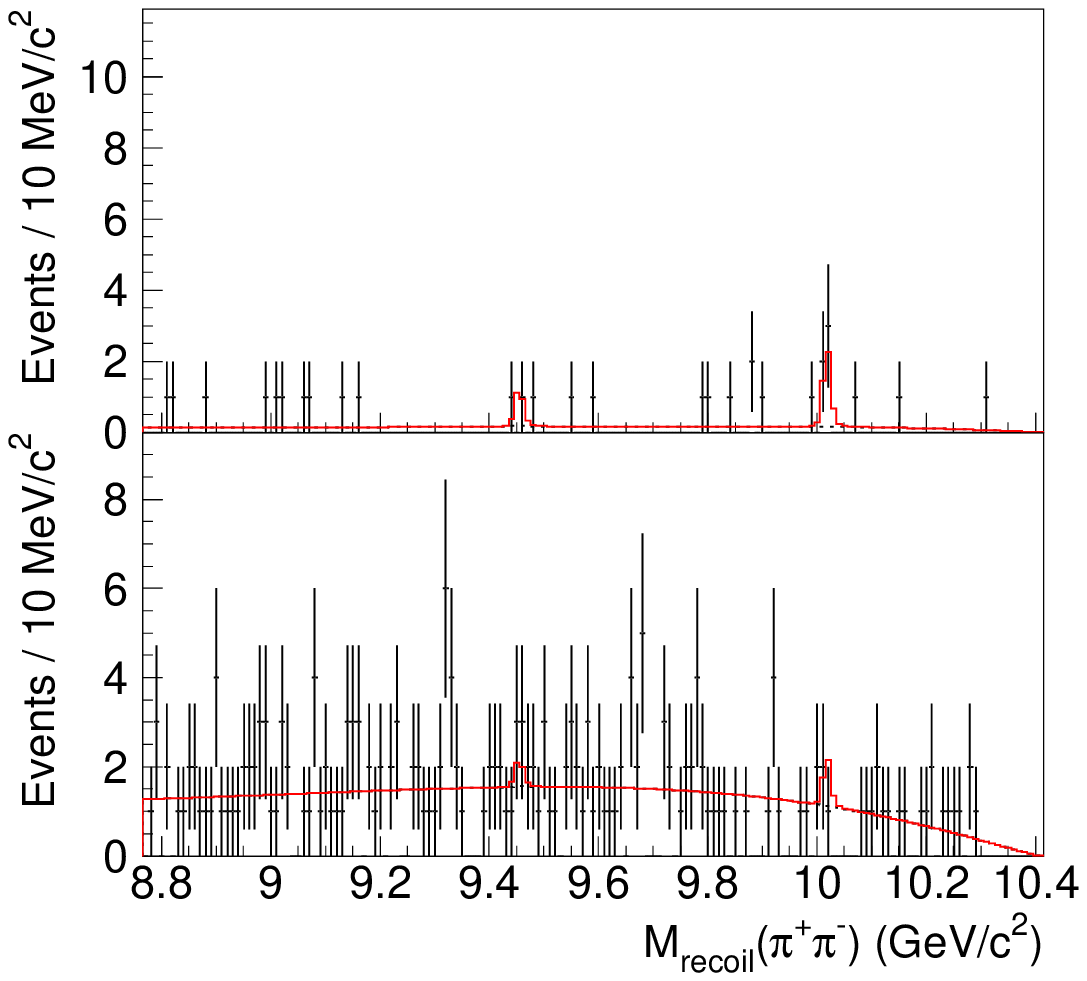}\hfill
\includegraphics[width=0.49\textwidth]{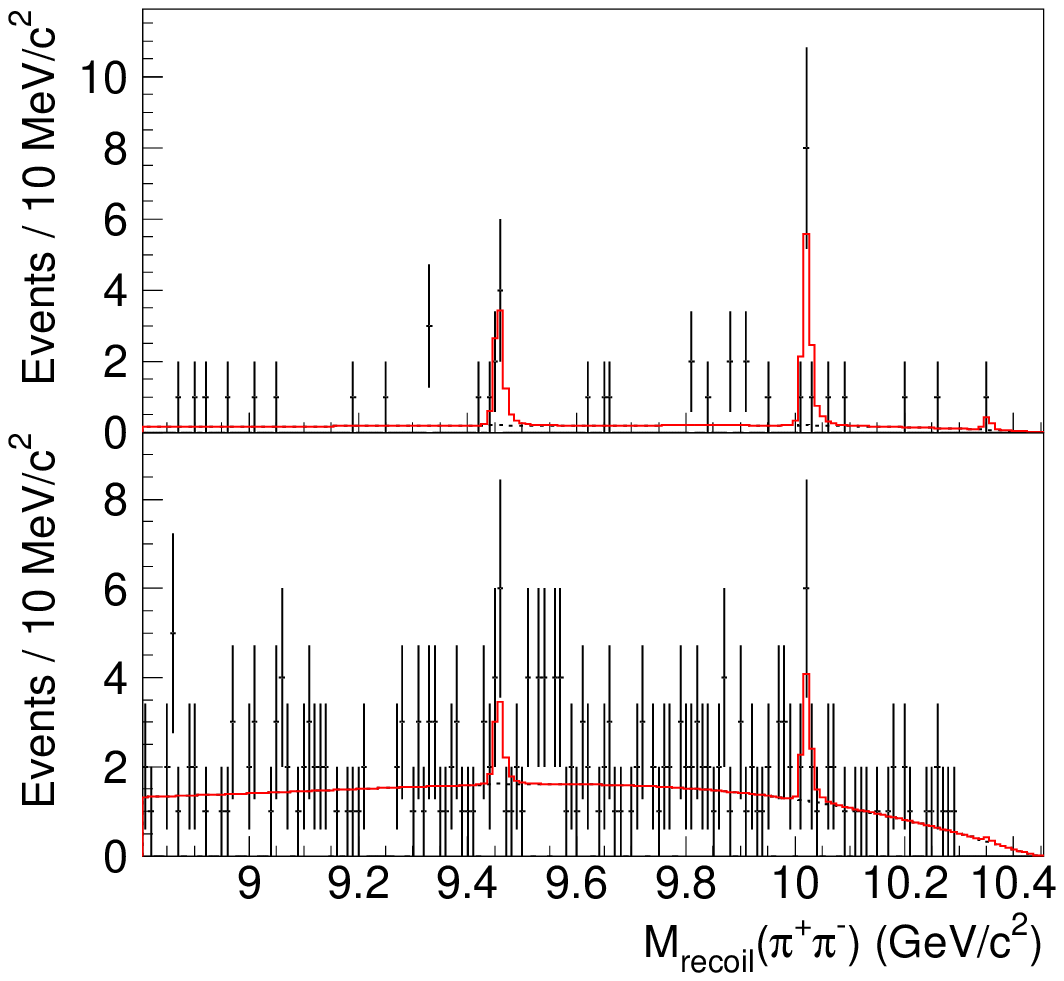}
\caption{ $\mmpp$ distributions in the data samples No.\ 4 (left) and
  5 (right); the numbering is given in Table~\ref{tab:ecm_lu_bo}. Top and
  bottom panels correspond to $\uu\pp$ and $\ee\pp$ final states,
  respectively.  Results of the fit (red solid histogram) and their
  background components (black dotted histogram) are also shown. }
\label{fit_uupp_eepp_24_101018}
\end{figure}

In case of the lowest energy, $10.52\,\gev$, reflections from the
$\Uth\to\Uo\pp$ and $\Uth\to\Ut\pp$ transitions leak into the signal
band in the 
$\muu$ vs.\ $\mmpp$ plane due to the finite resolution of the $\uu$
mass. The reflections are described in the fit by Gaussians with all
parameters floated.
Examples of the fits for the $\Uf$ on-resonance and scan data samples
are shown in Figs.~\ref{muu_vs_mmpp_exp69_051218} and
~\ref{fit_uupp_eepp_24_101018}.
The $\mmpp$ spectra are fitted with $1\,\mevm$ bins, but 
we present them with larger bin sizes for clarity.

\section{\boldmath Results for the center-of-mass energies and the 
$\Un\pp$ cross sections}
\label{sec:born}

The calibrated c.m.\ energies are obtained from the fit results for the
common shifts of the $\Un\pp$ signals. We do not find significant
deviations from the constraints of the $\ee\to\uu$ calibration; 
the $\ee\to\uu$ and $\ee\to\Un\pp$ energy calibrations agree
well. The calibrated $\ecm$ values are presented in
Table~\ref{tab:ecm_lu_bo}.
\begin{table}[!htp]
\renewcommand*{\arraystretch}{1.28}
\centering
\begin{tabular}{@{}cccccc@{}} \toprule
No. & $\ecm$ & Luminosity & $\sigma(\Uo\pp)$ & $\sigma(\Ut\pp)$ & $\sigma(\Uth\pp)$ \\ \midrule
  1 & $10517.1\pm1.4$ & 59.503 & $0.040^{+0.021}_{-0.019}$ & $0.025^{+0.029}_{-0.025}$ & $-$ \\
  2 & $10632.2\pm1.5$ &  0.989 & $0.21^{+0.23}_{-0.16}$ & $-0.08^{+0.19}_{-0.04}$ & $-$ \\
  3 & $10681.0\pm1.4$ &  0.949 & $-0.08^{+0.08}_{-0.03}$ & $0.06^{+0.27}_{-0.15}$ & $-$ \\
  4 & $10731.3\pm1.5$ &  0.946 & $0.34^{+0.33}_{-0.24}$ & $0.93^{+0.49}_{-0.38}$ & $-0.07^{+0.30}_{-0.08}$ \\
  5 & $10771.2\pm1.0$ &  0.955 & $1.08^{+0.42}_{-0.35}$ & $2.30^{+0.69}_{-0.59}$ & $0.29^{+0.47}_{-0.24}$ \\
  6 & $10829.5\pm1.2$ &  1.172 & $0.91^{+0.33}_{-0.28}$ & $1.89^{+0.55}_{-0.47}$ & $0.66^{+0.52}_{-0.38}$ \\
  7 & $10848.9\pm1.0$ &  0.989 & $2.08^{+0.51}_{-0.45}$ & $2.65^{+0.72}_{-0.62}$ & $1.39^{+0.66}_{-0.52}$ \\
  8 & $10857.4\pm0.9$ &  0.988 & $2.22^{+0.53}_{-0.47}$ & $3.17^{+0.81}_{-0.72}$ & $1.36^{+0.63}_{-0.50}$ \\
  9 & $10864.2\pm0.3$ & 47.647 & $1.957^{+0.083}_{-0.080}$ & $3.604^{+0.133}_{-0.131}$ & $1.181^{+0.088}_{-0.084}$ \\
 10 & $10865.8\pm0.3$ & 24.238 & $2.307^{+0.124}_{-0.121}$ & $4.306^{+0.205}_{-0.199}$ & $1.581^{+0.139}_{-0.131}$ \\
 11 & $10865.9\pm0.9$ &  1.814 & $1.94^{+0.38}_{-0.34}$ & $4.31^{+0.65}_{-0.60}$ & $1.08^{+0.41}_{-0.34}$ \\
 12 & $10867.5\pm0.4$ & 21.368 & $2.534^{+0.138}_{-0.133}$ & $3.949^{+0.211}_{-0.205}$ & $1.384^{+0.141}_{-0.133}$ \\
 13 & $10867.7\pm0.3$ & 22.938 & $2.287^{+0.128}_{-0.124}$ & $4.907^{+0.224}_{-0.218}$ & $1.608^{+0.144}_{-0.137}$ \\
 14 & $10867.9\pm0.8$ &  0.978 & $2.32^{+0.54}_{-0.48}$ & $5.29^{+0.96}_{-0.86}$ & $2.60^{+0.81}_{-0.68}$ \\
 15 & $10877.8\pm0.8$ &  0.978 & $2.86^{+0.60}_{-0.54}$ & $6.25^{+1.07}_{-0.97}$ & $1.94^{+0.73}_{-0.60}$ \\
 16 & $10882.8\pm0.7$ &  1.278 & $3.09^{+0.52}_{-0.47}$ & $7.91^{+1.01}_{-0.94}$ & $1.38^{+0.54}_{-0.45}$ \\
 17 & $10888.9\pm0.8$ &  0.990 & $1.93^{+0.48}_{-0.42}$ & $7.62^{+1.12}_{-1.04}$ & $2.02^{+0.69}_{-0.58}$ \\
 18 & $10898.3\pm0.9$ &  0.983 & $1.65^{+0.42}_{-0.37}$ & $5.70^{+0.93}_{-0.85}$ & $1.48^{+0.55}_{-0.45}$ \\
 19 & $10898.3\pm1.0$ &  0.883 & $2.13^{+0.50}_{-0.44}$ & $5.38^{+0.96}_{-0.87}$ & $1.12^{+0.55}_{-0.43}$ \\
 20 & $10907.3\pm1.1$ &  0.980 & $1.17^{+0.34}_{-0.29}$ & $2.38^{+0.60}_{-0.53}$ & $0.48^{+0.35}_{-0.25}$ \\
 21 & $10928.7\pm1.6$ &  0.680 & $0.35^{+0.24}_{-0.19}$ & $1.71^{+0.65}_{-0.52}$ & $0.52^{+0.35}_{-0.26}$ \\
 22 & $10957.5\pm1.5$ &  0.855 & $0.10^{+0.15}_{-0.10}$ & $0.86^{+0.42}_{-0.35}$ & $0.22^{+0.30}_{-0.22}$ \\
 23 & $10975.3\pm1.4$ &  0.999 & $0.76^{+0.37}_{-0.31}$ & $1.00^{+0.43}_{-0.35}$ & $0.41^{+0.31}_{-0.23}$ \\
 24 & $10990.4\pm1.3$ &  0.985 & $1.60^{+0.49}_{-0.42}$ & $1.86^{+0.58}_{-0.49}$ & $0.48^{+0.38}_{-0.27}$ \\
 25 & $11003.9\pm1.0$ &  0.976 & $1.71^{+0.46}_{-0.41}$ & $2.15^{+0.62}_{-0.54}$ & $1.49^{+0.57}_{-0.49}$ \\
 26 & $11014.8\pm1.4$ &  0.771 & $0.88^{+0.37}_{-0.31}$ & $1.58^{+0.64}_{-0.54}$ & $0.87^{+0.45}_{-0.35}$ \\
 27 & $11018.5\pm2.0$ &  0.859 & $0.56^{+0.29}_{-0.23}$ & $1.21^{+0.50}_{-0.42}$ & $0.33^{+0.32}_{-0.23}$ \\
 28 & $11020.8\pm1.4$ &  0.982 & $0.61^{+0.28}_{-0.23}$ & $1.02^{+0.45}_{-0.38}$ & $0.61^{+0.36}_{-0.28}$ \\
\bottomrule
\end{tabular}
\caption{ Calibrated c.m.\ energy (in GeV), luminosity of various data
  samples (in $\fb$) and measured Born cross sections of the
  $\ee\to\Un\pp$ ($n=1,2,3$) processes (in pb). The uncertainties in
  the energies and cross sections are totally uncorrelated. }
\label{tab:ecm_lu_bo}
\end{table}

\begin{figure}[htbp]
\centering
\includegraphics[width=0.43\textwidth]{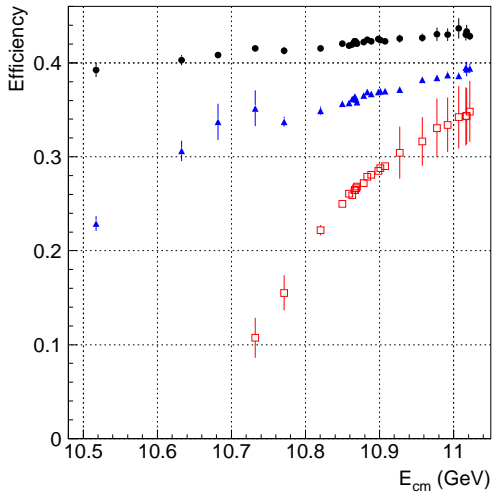}
\caption{ Reconstruction efficiency as a function of the c.m.\ energy
  for the $\Uo\pp$ (black filled dots), $\Ut\pp$ (blue triangles), and
  $\Uth\pp$ (red open dots) channels. The error bars show the
  systematic uncertainties associated with the DP model. }
\label{eff_vs_ecm_050918}
\end{figure}

Based on the measured signal yields we calculate Born cross sections:
\begin{equation}
  \sigma^B=\frac{N\; |1-\Pi|^2}{\varepsilon\; L\; {\cal B}},
  \label{eq:sigma_born}
\end{equation}
where $N$ is the signal yield (since the signal density function has
been appropriately normalized, $N$ includes the ISR correction, as
discussed in Section~\ref{sec:shape}), $|1-\Pi|^2$ is a vacuum
polarization correction taken from Ref.~\cite{Actis:2010gg},
$\varepsilon$ is the reconstruction efficiency
(Fig.~\ref{eff_vs_ecm_050918}), $L$ is the integrated luminosity of
each data sample (Table~\ref{tab:ecm_lu_bo}), and ${\cal B}$ is the
branching fraction for $\Un\to\uu$~\cite{Patrignani:2016xqp}.
Measured $\ee\to\Un\pp$ cross sections are presented in
Table~\ref{tab:ecm_lu_bo}.

To study systematic uncertainties in $\ecm$ and the cross sections, we
vary the $\ecm$ spread, the scale factor $f$, and the $\ee\pp$ to $\uu\pp$
efficiency ratio by $\pm1\,\sigma$. We also increase the polynomial
order in the background parameterization for high-luminosity data
samples, while for low luminosity we release the
coefficient of the linear term of the Chebyshev polynomial.
Uncertainties associated with the cross section energy dependence are
estimated using MC pseudoexperiments that are generated according to
the fit results described in Section~\ref{sec:fit}. The signal density
functions are computed based on the fit results of each
pseudoexperiment.

The systematic uncertainties are estimated based on the deviations of
the measured quantities from their nominal values under the above
variations of the analysis. In order to avoid overestimation of the
relative uncertainties among the $\Uf$ on-resonance points, we
separate uncertainties into correlated and uncorrelated parts. For
each variation, we first find the average deviation over all the
on-resonance points; this is used to estimate the correlated
uncertainty. Then, we subtract the average deviation from deviations
at each energy point. These relative deviations are used to estimate
the uncorrelated uncertainties. In the case of the cross section
energy dependence, we take the root mean squares of the deviations
(for both correlated and uncorrelated parts). In all other cases, we
take the maximal deviation as the uncertainty.

\begin{figure}[htbp]
\centering
\includegraphics[width=0.43\textwidth]{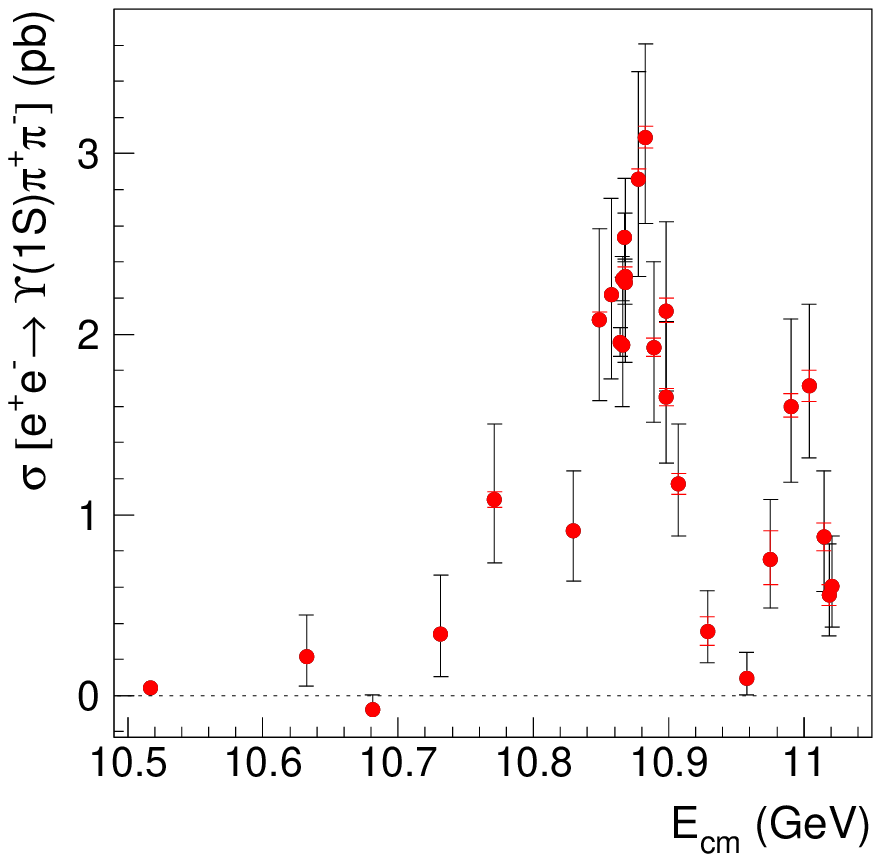}\hfill
\includegraphics[width=0.43\textwidth]{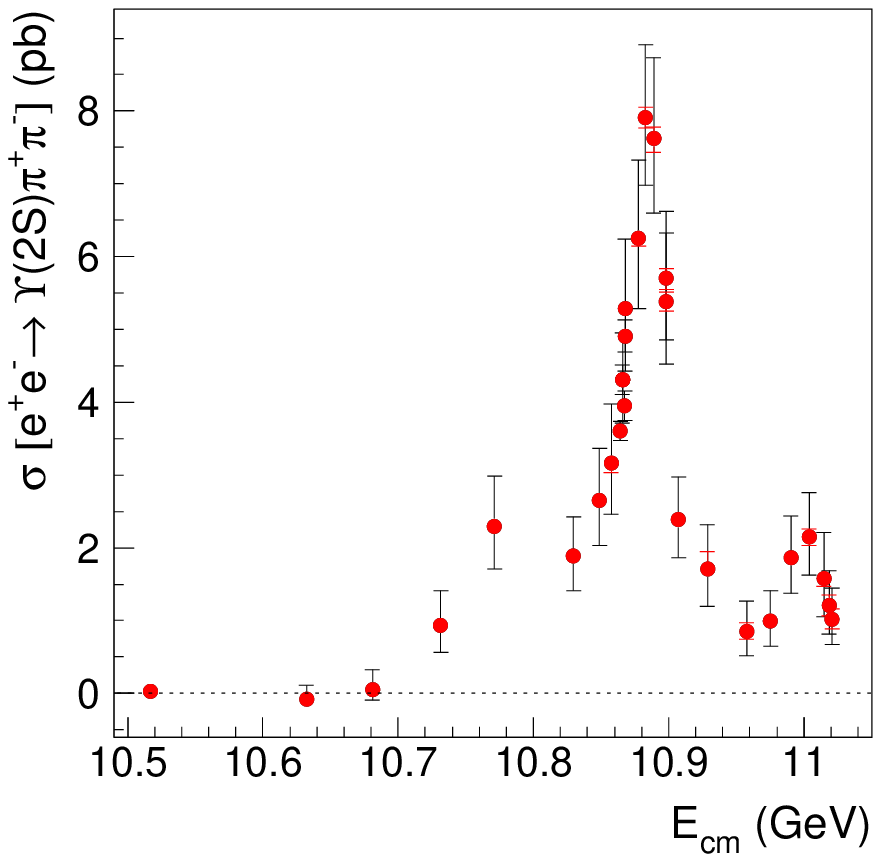}\hfill
\includegraphics[width=0.43\textwidth]{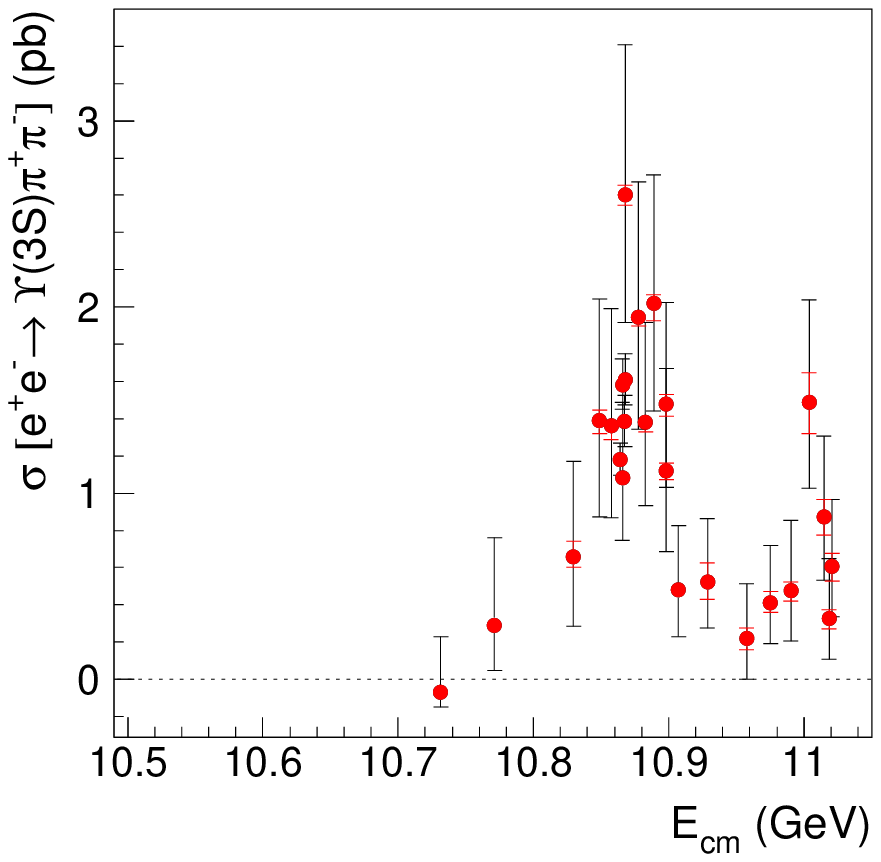}
\caption{ Born cross sections for $\ee\to\Un\pp$ ($n=1,2,3$) with
  statistical (black) and uncorrelated systematic (red)
  uncertainties, measured for different c.m.\ energies. }
\label{xsec_no_fit}
\end{figure}

Uncertainties in efficiency associated with the DP model are
treated as uncorrelated.
The total uncorrelated systematic uncertainty is the sum in quadrature
of all the contributions; dominant among them is the cross section
energy dependence. The systematic uncertainty is small compared to the
statistical uncertainty, as can be seen in Fig.~\ref{xsec_no_fit}. We
add the two contributions in quadrature to find the total uncorrelated
uncertainty.
The c.m.\ energies and Born cross sections
with total uncorrelated uncertainties are presented in
Table~\ref{tab:ecm_lu_bo}.

The correlated systematic errors include the uncertainties in the
efficiency due to possible differences between data and MC in track
reconstruction (0.35\% per high and 1\% per low momentum tracks, which
are muons and pions, respectively) and muon identification, and the
uncertainties in the luminosity and the branching fractions for
$\Un\to\uu$ decays. The summary of the correlated errors is presented
in Table~\ref{tab:correlated_syst}.
\begin{table}[htbp]
\caption{ The systematic uncertainties in the $\ee\to\Un\pp$
  ($n=1,2,3$) cross sections correlated among various energy points
  (in percent). }
\label{tab:correlated_syst}
\centering
\begin{tabular}{@{}lccc@{}} \toprule
& $\Uo\pp$ & $\Ut\pp$ & $\Uth\pp$ \\ \midrule
$\ecm$ spread & 0.2 & 0.2 & 0.4 \\
Momentum resolution & 0.3 & 0.1 & 0.1 \\
Cross section energy dependence & 0.4 & 0.9 & 0.8 \\
Tracking & 2.7 & 2.7 & 2.7 \\
Muon identification & 2 & 2 & 2 \\
Luminosity & 1.4 & 1.4 & 1.4 \\
Branching fractions & 2.0 & 8.8 & 9.6 \\ \midrule
Total & 4.2 & 9.6 & 10.3 \\
\bottomrule
\end{tabular}
\end{table}

The energy-dependent cross sections (Fig.~\ref{xsec_no_fit}) show
clear $\Uf$ and $\Us$ peaks that were seen in previous
publications~\cite{Chen:2008xia,Santel:2015qga}.
Due to the improved precision an additional structure around
$\ecm=10.75\,\gev$, hinted at by the measurements in
Ref.~\cite{Santel:2015qga}, is visible.
The significance of this state is quantified in Section~\ref{sec:fit}.

Recently, Belle studied the $\ee\to\Un\pp$ ($n=1,2$) processes in the
$\U(4S)$ on-resonance data~\cite{Guido:2017cts} and reported visible
cross sections. For completeness, we include the corresponding Born
cross sections here. We determine the $(1+\delta_{\mathrm{ISR}})$
factor to be $0.611\pm0.011$, where the error includes uncertainty in
the $\U(4S)$ parameters and uncertainty in the ISR photon energy
cut-off. The Born cross sections at the $\U(4S)$ peak, recalculated
based on the results of Ref.~\cite{Guido:2017cts}, are
\begin{align}
  \sigma(\ee\to\Uo\pp) & = (0.1350\pm0.0080\pm0.0071)\,\mathrm{pb},\\
  \sigma(\ee\to\Ut\pp) & = (0.1293\pm0.0134\pm0.0132)\,\mathrm{pb}.
  \label{eq:xsec_y4s}
\end{align}

\section{\boldmath Study of the continuum data sample} 
\label{sec:conti}

In the continuum data sample at $\ecm=10.52\,\gev$, we find hints of
non-zero values for the $\Uo\pp$ and $\Ut\pp$ cross sections,
$40^{+21}_{-19}\,$fb and $25^{+29}_{-25}\,$fb, respectively.
At this energy, contributions to the $\Un\pp$ cross sections due to
$\U(4S)$, $\Uf$, or the new structure at $10.75\,\gev$ are negligible
(see next Section).
To estimate the contributions of the $\Ut$ and $\Uth$ tails we use the
Breit-Wigner function of Eq.~(\ref{eq:bw}) and world-average values
for the $\U(1S,2S)$ masses, widths, electron widths and branching
fractions~\cite{Patrignani:2016xqp}.
We find 71\,fb, 2\,fb and 35\,fb for the $\Ut\to\Uo\pp$,
$\Uth\to\Uo\pp$, and $\Uth\to\Ut\pp$ tails, respectively, which values
are close to the experimental measurements.

\begin{figure}[htbp]
\centering
\includegraphics[width=0.43\textwidth]{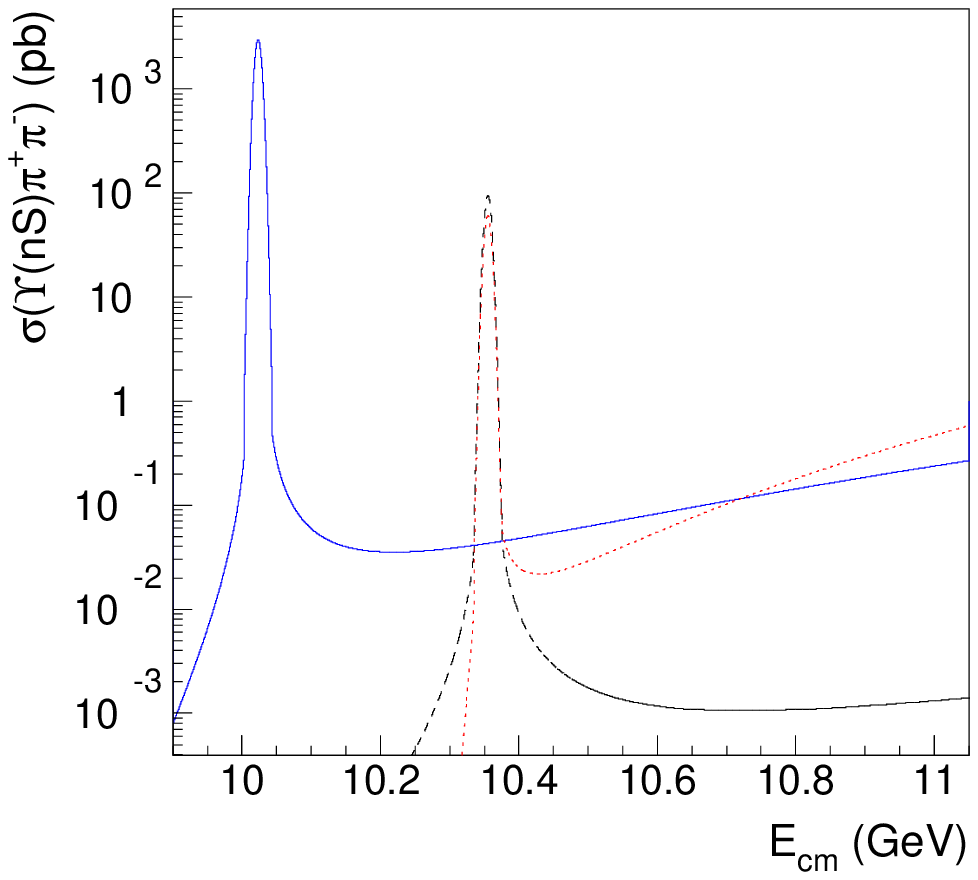}
\caption{ Expected energy dependence of the cross sections for the
  processes $\ee\to\Ut\to\Uo\pp$ (solid blue), $\ee\to\Uth\to\Uo\pp$
  (dashed black) and $\ee\to\Uth\to\Ut\pp$ (dotted red). These include
  the c.m.\ energy spread of $\sigma=5\,\mev$. }
\label{tails_vs_ecm_200319}
\end{figure}

The obtained expectations for the $\Ut\to\Uo\pp$ and $\Uth\to\Ut\pp$
tails are quite large because matrix elements of these three-body
decays are dominated by the terms proportional to $\mppsq$ (see,
e.g.,~\cite{CroninHennessy:2007zz}).
Corresponding partial widths are calculated as integrals of the matrix
elements over phase space; they increase rapidly with the c.m.\ energy
as high values of $\mpp$ become kinematically allowed.
The dependence of the cross sections for various tails on $\ecm$ is
shown in Fig.~\ref{tails_vs_ecm_200319}. 
The cross sections start to increase above a certain energy. 
Uncertainties of the prediction of the tail contributions are
discussed in the next Section.

\begin{figure}[htbp]
\centering
\includegraphics[width=0.43\textwidth]{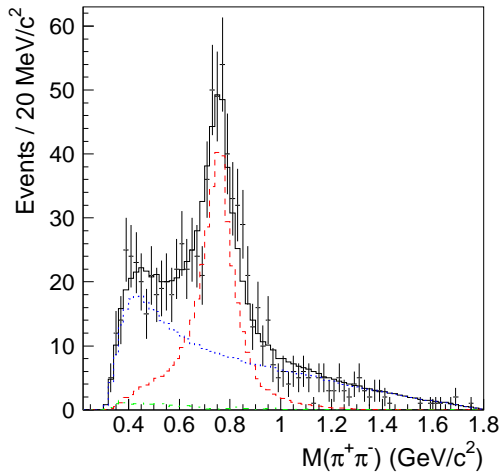}
\caption{ The $\mpp$ distribution for the $\uu\pp$ events in the $\Uo$
  sideband region $8.7<\mmpp<9.4\,\gevm$ at $\ecm=10.52\,\gev$. Points
  with error bars show data, the histograms show MC simulations of the
  total QED background (solid black) and various background
  components: $\uu\pp$ (dashed red), misidentified $4\mu$ (dotted
  blue), and misidentified $\ee\uu$ (dash-dotted green). }
\label{mpp_data_mc_87_94_130219}
\end{figure}

\begin{figure}[htbp]
\centering
\includegraphics[width=0.43\textwidth]{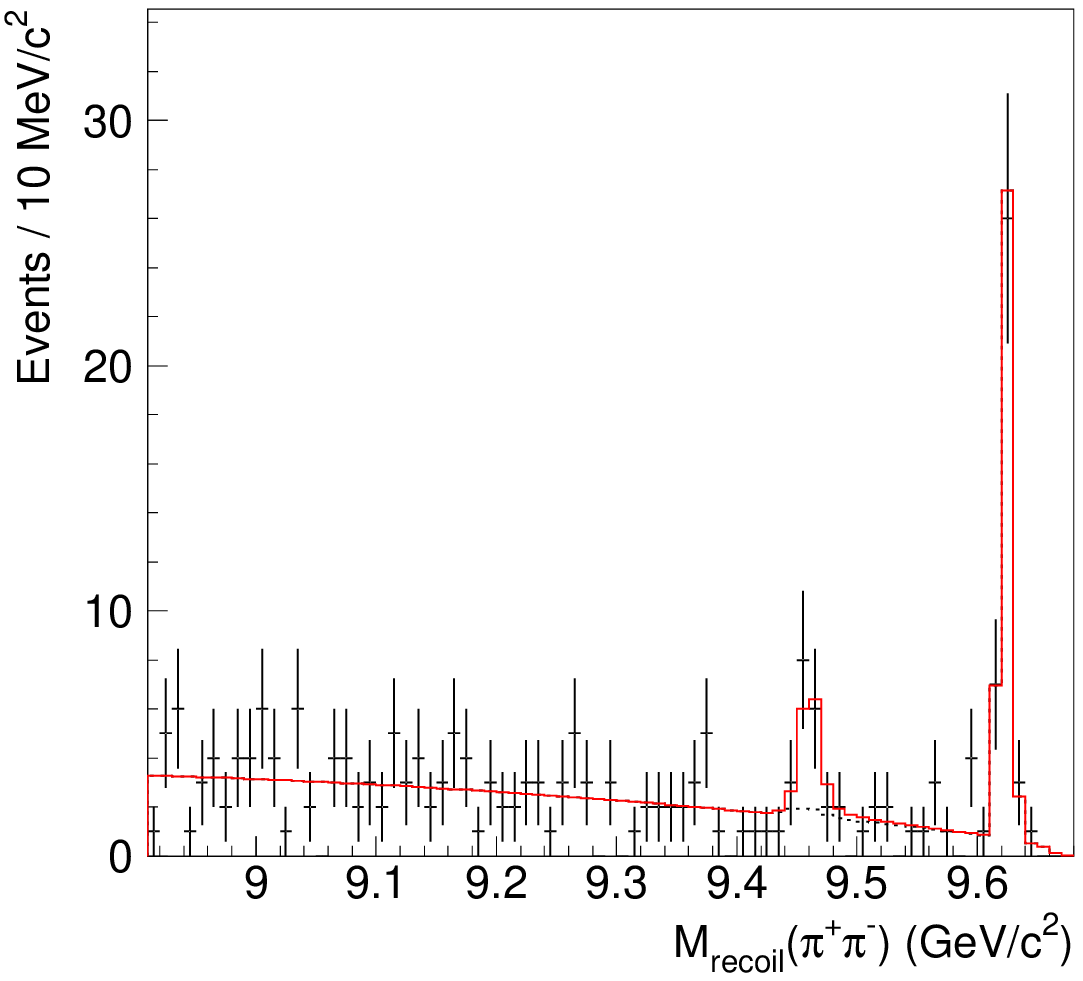}
\caption{ The $\mmpp$ distribution for $\uu\pp$ events with an
  additional requirement of $\mpp>0.85\,\gevm$ for the data sample at
  $\ecm=10.52\,\gev$. The peak at $9.63\,\gevm$ is due to the
  $\Ut\to\Uo\pp$ transitions with ISR production of the $\Ut$. Result
  of the fit (red solid histogram) and the background component (black
  dotted histogram) are also shown. }
\label{mmpp_conti_uupp_fit_111218}
\end{figure}

To verify the hypothesis of the $\U(2S,3S)$ tails, we note that the
$\mpp$ distribution for the QED background is quite different from the
expectations for the tails. The background is studied using the
$\uu\pp$ events in the $\Uo$ sideband region $8.7<\mmpp<9.4\,\gevm$.
The background $\mpp$ distribution
(Fig.~\ref{mpp_data_mc_87_94_130219}) is enhanced at low values and in
the region of the $\rho$ meson; the MC simulation describes the shape
and the normalization of the sideband data quite well.
In contrast, the $\U(2S,3S)$ tail events are expected to concentrate
near the upper kinematic boundary.
To suppress the QED background, we apply a requirement
$\mpp>0.85\,\gevm$, which is optimized using the figure of merit
$\frac{S}{3/2+\sqrt{B}}$~\cite{Punzi:2004wh}, where $S$ and $B$ are
the number of signal and background events in the simulation,
respectively.
The $\pp$ recoil mass distribution with this additional requirement
for the $\uu\pp$ final state is shown in
Fig.~\ref{mmpp_conti_uupp_fit_111218}.
A clear signal for the $\Uo\pp$ process is evident; its significance
is estimated using Wilks' theorem~\cite{wilks} to be
$3.6\,\sigma$. The signal is stable against variations of the fit
interval and the order of the polynomial used for the background
parameterization.
We conclude that the significance including systematic uncertainties
is larger than $3.5\,\sigma$. 
The expected $\mpp$ requirement efficiency is 72\% and is consistent
with the reduction of the signal yield in the data.
Based on the signal yield measured with the $\mpp$ requirement, we
determine the $\ee\to\Un\pp$ cross section at $\ecm=10.52\,\gev$ to be
$0.042^{+0.017}_{-0.015}$\,pb, where the uncertainties are statistical
only. This result is model-dependent because of the unknown $\mpp$
distribution. In the fit to the cross section energy dependence, we
use the value measured without the $\mpp$ requirement.

\section{Fit to the energy dependence of the cross sections}
\label{sec:fit}

Since the numbers of signal events in some scan data samples are
small, the errors in the measured cross sections might be
non-Gaussian. To address this issue, we scan $-2\ln\lik$ of the $\mmpp$
fits in the signal yields. The $-2\ln\lik$ dependence on the signal
yield is then converted into the $-2\ln\lik$ dependence on the cross
section and is parameterized using an empirical function
\begin{equation}
  f(x) = 2 \cdot (x\cdot p_2 + p_3 - p_1 +
  p_1\ln\frac{p_1}{x\cdot p_2 + p_3}) \cdot P_6(x),
\end{equation}
where $p_1$, $p_2$ and $p_3$ are fit parameters and $P_6$ is a 
6th-order Chebyshev polynomial, with the zeroth-order parameter set to 
unity and all other parameters floated.
To account for the systematic uncertainty, we convolve the
distributions in $\lik$ with Gaussian functions that represent
uncorrelated systematic uncertainties.
We find that there is good agreement between the scan results and
the asymmetric Gaussian errors in the $\pm1\,\sigma$ region; however,
there are noticeable discrepancies for larger deviations. Therefore
Gaussian errors should give a good approximation for a default fit that
describes data well; while the $-2\ln\lik$ scan results are important
for significance calculation, in which case some points deviate from
the alternative fit function by several standard deviations.
In the fits to the cross section energy dependence, we use the
$-2\ln\lik$ scan results for all scan data points outside the $\Uf$
peak (points 2 to 8 and 20 to 28 in Table~\ref{tab:ecm_lu_bo}).

The ISR tails of the $\Un\pp$ signals in the $\Uf$ on-resonance data
are sensitive to the cross section shapes in the region of the new
structure at $10.75\,\gev$. Therefore we include the $\mmpp$ distribution for the
$\uu\pp$ final state, where background is lower, into the fit of the
cross section energy dependence. Thus we perform a simultaneous fit to
the $\Un\pp$ ($n=1,2,3$) cross sections and the $\mmpp$ distribution.

To parameterize the cross sections energy dependence, we consider
contributions from the $\Uf$ and $\Us$ resonances, the new structure,
and the $\U(2S,3S)$ tails. Each contribution is represented by a
Breit-Wigner amplitude:
\begin{equation}
F_{BW}(s,M,\Gamma,\Gamma_{ee}^0\times\br_f) =
    \frac{\sqrt{12\pi\;\Gamma\;\Gamma_{ee}^0\times\br_f}}{s-M^2+iM\Gamma}\,
    \sqrt{\frac{\Gamma_f(s)}{\Gamma_f(M^2)}},
    \label{eq:bw}
\end{equation}
where $M$ and $\Gamma$ are the mass and total width, $\Gamma_{ee}^0$
is the ``bare'' electron width related to the physical width by
$\Gamma_{ee}^0=\Gamma_{ee}\,|1-\Pi(s)|^2$, and $\br_f$ and
$\Gamma_f(s)$ are branching fraction and energy-dependent partial
width of the decay to the $\Un\pp$ final state.
The values of $\Gamma_f(s)$ at various energies are computed
numerically by integrating the three-body decay matrix element over
the DPs.

The new structure might have either resonant or non-resonant origin.
In some cases, non-resonant effects produce peaks and have phase
motion similar to resonances, as discussed, e.g., in
Ref.~\cite{Bugg:2011jr}. Thus, we use the Breit-Wigner
parameterization for the new structure amplitude. 

In the default model the DP distributions of the $\Uf$, $\Us$, and new
structure decays are assumed to be the same, thus their amplitudes are
added coherently. For the $\Ut\to\Uo\pp$ and $\Uth\to\Ut\pp$ tails we
use the three-body matrix elements measured by
CLEO~\cite{CroninHennessy:2007zz}. The interference terms between the
tails and the rest of the amplitude are multiplied by ``decoherence
factors'' that are calculated as overlap integrals of the DP matrix
elements~\cite{Santel:2015qga,Abdesselam:2015zza}, and can take values
that range from 0 (incoherence) to 1 (full coherence). The values of
the decoherence factors are found to be typically 0.7--0.8.

Complex phases of the $\Us$, the new structure, and the $\U(2S,3S)$
amplitudes relative to the $\Uf$ amplitude are all floated in the fit
individually for the three channels. We also float the $M$, $\Gamma$
and $\Gamma_{ee}^0\times\br_f$ parameters of the Breit-Wigner
amplitudes except for $M$ and $\Gamma$ of the $\Ut$ and $\Uth$.
The fit function to the $\Un\pp$ cross section at each c.m.\ energy
contains a convolution with a Gaussian to account for the energy
spread.
The fit function to the $\mmpp$ distribution is the same as described
in Section~\ref{sec:mmpp_fits}.

\begin{figure}[htbp]
\centering
\includegraphics[width=0.67\textwidth]{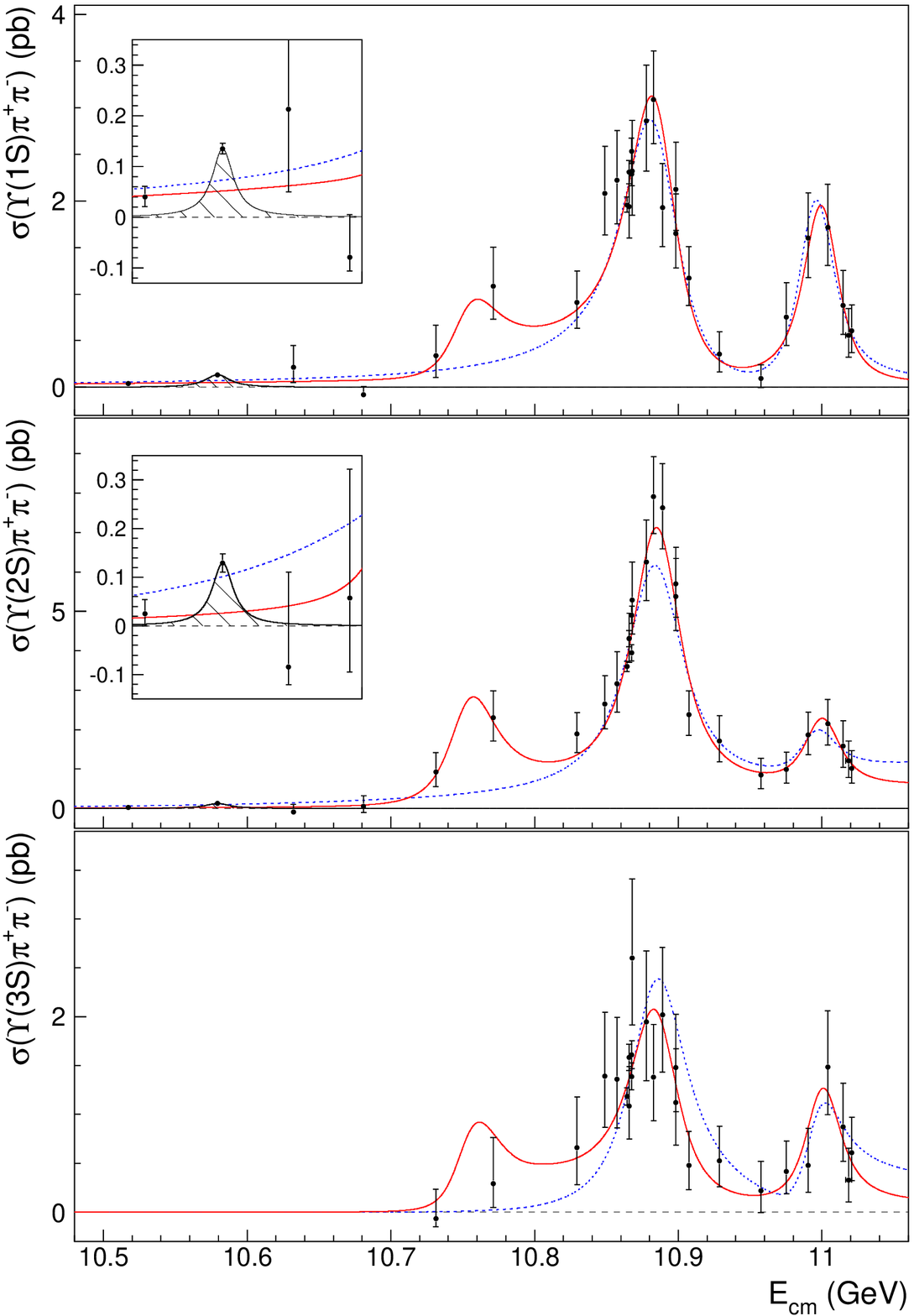}\hfill
\includegraphics[width=0.2931\textwidth]{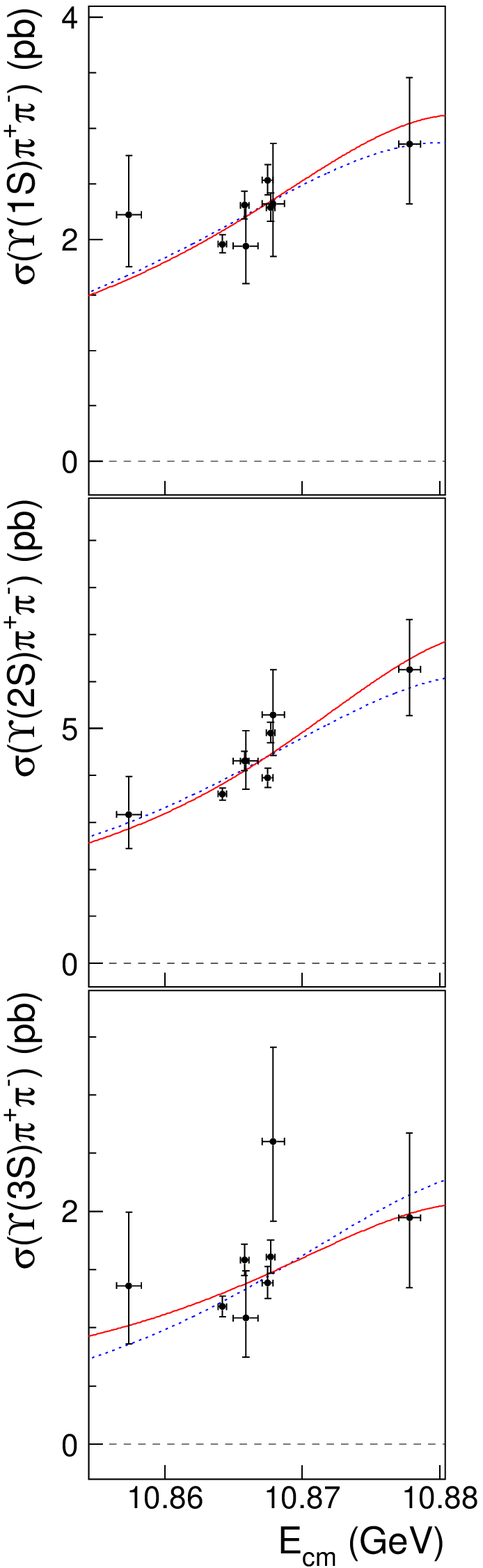}
\caption{ Energy dependence of the $\ee\to\Un\pp$ cross sections
  ($n=1,2,3$ from top to bottom). The points with error bars are data
  with combined statistical and uncorrelated systematic uncertainties;
  the curves are the results of the simultaneous fit to these
  distributions and the $\mmpp$ distribution in the on-resonance data
  (Fig.~\ref{fitted_mmpp_3model_031118}). Shown are the results of
  the default fit (red solid curve) and the fit with the new structure
  excluded in all channels (blue dotted curve). The hatched histograms
  and the points with error bars at their maxima show the $\U(4S)$
  line shapes and the cross sections measured at the $\U(4S)$ peak in
  Ref.~\cite{Guido:2017cts}. These points are not used in the fit to
  the cross section energy dependence. Insets show a zoom of the low
  energy region, while right side panels show a zoom of the $\Uf$
  on-resonance region. }
\label{fitted_xsec_3model_031118}
\end{figure}

\begin{figure}[htbp]
\centering
\includegraphics[width=0.49\textwidth]{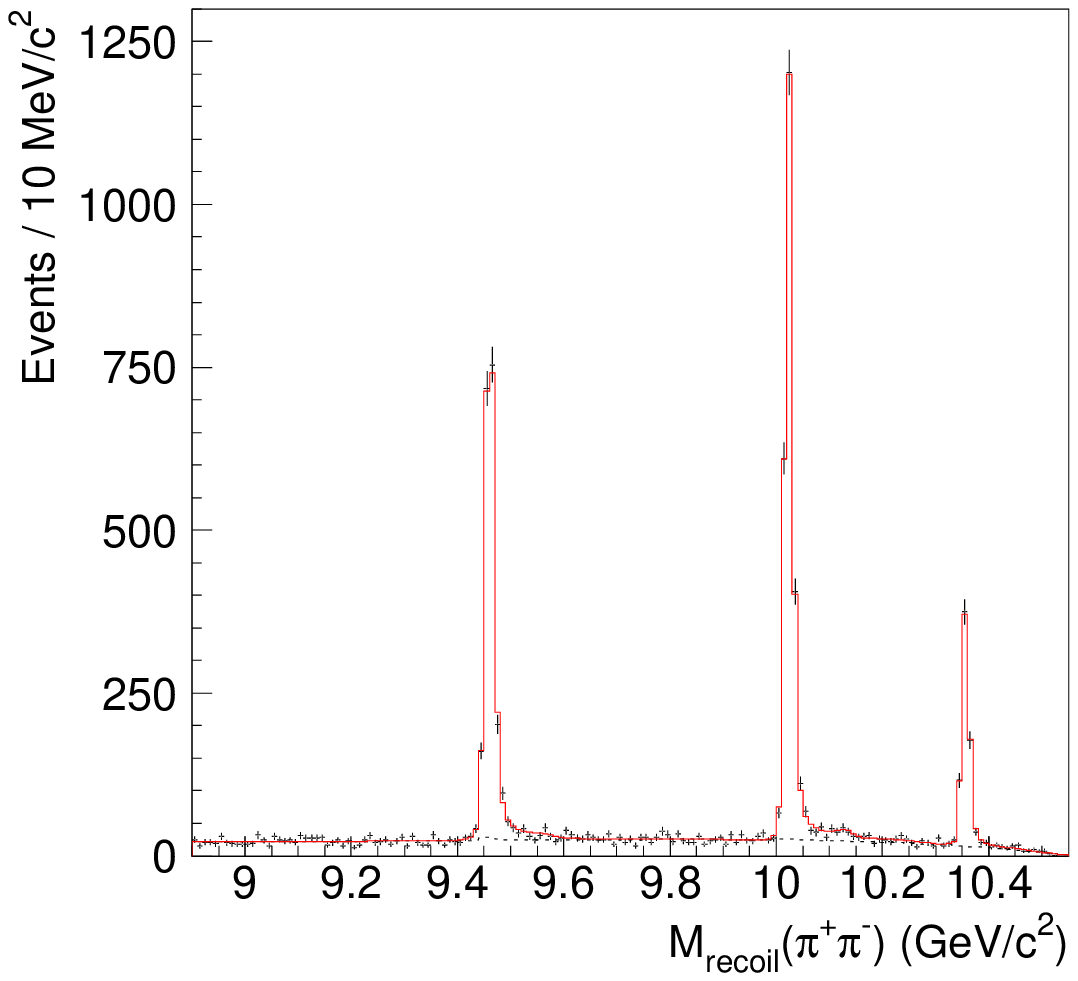}\hfill
\includegraphics[width=0.49\textwidth]{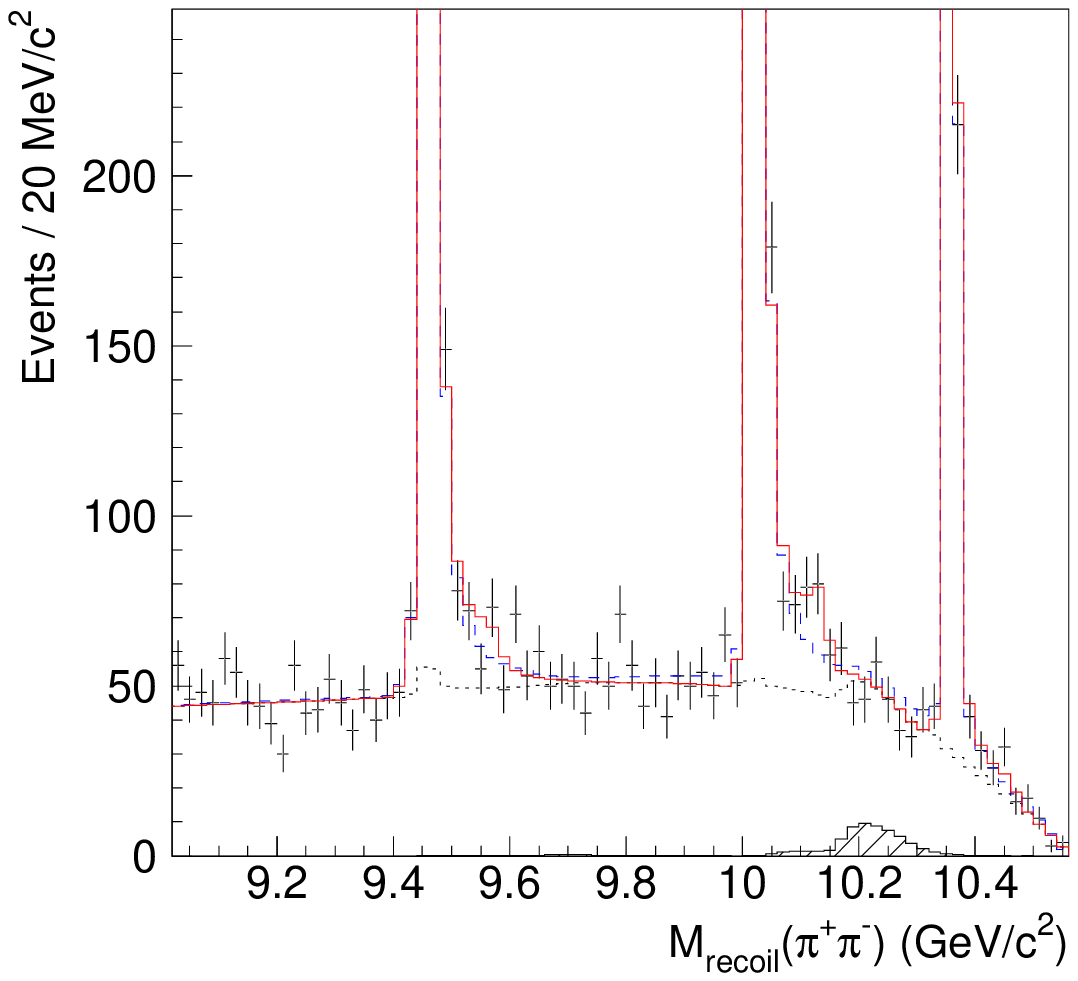}
\caption{ The $\mmpp$ distribution for the $\uu\pp$ events in the
  $\Uf$ on-resonance data; the right panel is a zoom of the left
  panel. Histograms show the results of the simultaneous fit to this
  distribution and the energy dependence of the $\ee\to\Un\pp$
  ($n=1,2,3$) cross sections
  (Fig.~\ref{fitted_xsec_3model_031118}). The red solid histogram
  corresponds to the default fit, while the blue dashed histogram in
  the right panel corresponds to the fit with the new structure
  excluded.  The black dotted histogram shows the background component
  of the default fit, the hatched one -- peaking background from
  $\eta$ decays. }
\label{fitted_mmpp_3model_031118}
\end{figure}

The fit results are presented in Figs.~\ref{fitted_xsec_3model_031118}
and \ref{fitted_mmpp_3model_031118}.
For illustration, we show in Fig.~\ref{fitted_xsec_3model_031118} the
cross sections measured at the $\U(4S)$ peak and the expected $\U(4S)$
line shape; these measurements are not used in the fit. 
The fit results for masses and widths of the $\Uf$, $\Us$, and new
structure are presented in Table~\ref{tab:m_g_stat_syst}.
\begin{table}[htbp]
\caption{ Measured masses and widths of the $\Uf$, $\Us$ and the new
  structure. The first uncertainty is statistical, the second is
  systematic. }
\renewcommand*{\arraystretch}{1.5}
\label{tab:m_g_stat_syst}
\centering
\begin{tabular}{@{}clll@{}} \toprule
& $\Uf$ & $\Us$ & New structure \\ \midrule
$\rm M\;(\mevm)$ & $10885.3\pm1.5\,^{+2.2}_{-0.9}$   & $11000.0^{+4.0}_{-4.5}\,^{+1.0}_{-1.3}$ & $10752.7\pm5.9\,^{+0.7}_{-1.1}$ \\
$\Gamma\;(\mev)$ & $36.6^{+4.5}_{-3.9}\,^{+0.5}_{-1.1}$ & $23.8^{+8.0}_{-6.8}\,^{+0.7}_{-1.8}$    & $35.5^{+17.6}_{-11.3}\,^{+3.9}_{-3.3}$ \\
\bottomrule
\end{tabular}
\end{table}
The results are slightly shifted from the previous Belle measurements
based on the same data sample~\cite{Santel:2015qga} because we updated
the cross section values at each energy, we fit now Born cross
sections instead of visible ones, the new structure and the
non-resonant contributions, i.e. the $\U(2S,3S)$ tails, are included
in the fit model, and the energy dependence of the $\Gamma_f$ is taken
into account.

A sum of several Breit-Wigner amplitudes produces multiple solutions
that have the same values of $-2\ln L$, the same masses and widths,
but different normalizations and relative phases. To search for
multiple solutions, we create points with fitted values of the cross
sections and small uncertainties. We then fit these points, one
$\Un\pp$ channel at a time, repeating the fit many times with randomly
generated initial values of the fit parameters. We find four or eight
solutions in various channels, as expected for the sum of three or
four Breit-Wigner amplitudes, respectively. The $\pm1\,\sigma$
intervals of various solutions typically overlap, therefore we present
only ranges in $\gee\times\br_f$ from the lowest to the highest
solution (Table~\ref{tab:intervals}).
\begin{table}[htbp]
\caption{ The ranges of the $\gee\times\br$ values from the multiple
  solutions (in $\ev$). }
\renewcommand*{\arraystretch}{1.1}
\label{tab:intervals}
\centering
\begin{tabular}{@{}l|ccc@{}} \toprule
          & $\Uf$       & $\Us$       & new \\ \midrule
$\Uo\pp$  & $0.75-1.43$ & $0.38-0.54$ & $0.12-0.47$ \\
$\Ut\pp$  & $1.35-3.80$ & $0.13-1.16$ & $0.53-1.22$ \\
$\Uth\pp$ & $0.43-1.03$ & $0.17-0.49$ & $0.21-0.26$ \\
\bottomrule
\end{tabular}
\end{table}

To estimate the significance of the new structure in a single $\Un\pp$
channel, we repeat the fit with the new structure excluded in that
channel. The significance is found based on the change in the $-2\ln
L$ of the fit using Wilks' theorem. The values are 2.7 and 5.4
standard deviations (${\sigma}$) in the channels $\Uo\pp$ and
$\Ut\pp$, respectively. Using large sample of MC pseudoexperiments, we
find that Wilks' theorem provides slightly conservative estimation in
our case. This is related to the fact that the number of experimental
points in the region of the new structure is quite limited.

We also estimate the significance for all three $\Un\pp$ channels
combined by performing the fit with the new structure excluded in all
channels simultaneously. The $-2\ln{L}$ difference between the default
fit and the null hypothesis fit is 65.8, with the cross section scan
results contributing 51.7 and recoil mass distribution contributing
14.1. This $-2\ln{L}$ difference corresponds to a local significance of
$7.0\,\sigma$, estimated using Wilks' theorem. The global
significance is estimated using the method described in
Ref.~\cite{Vitells:2011da}. We find that the p-value of the
fluctuation increases by a factor 4.5 due to the ``look-elsewhere
effect'', and the resulting global significance is $6.8\,\sigma$.

To estimate systematic uncertainties in the $\Uf$, $\Us$, and new
structure parameters, we vary the fit procedure.
As an alternative parameterization of the new structure amplitude we
use the threshold function, Eq.~(\ref{eq:argus}). For its parameters
we find $x_0\approx10.73\,\gevm$, $p\approx0.17$ and
$c\approx-10\,(\gevm)^{-1}$. The quality of this fit is comparable to
the default fit, with $-2\ln L$ worse by 3.4. Such a parameterization
represents a threshold-like contribution without variation of a
complex phase with energy. It has no clear physical meaning and we use
it to conservatively estimate systematic uncertainty in the $\Uf$ and
$\Us$ parameters.

We study the influence of thresholds on the line shape of the $\Uf$
resonance. We consider the thresholds in the region of the new
structure that show strong coupling to the $\Uf$. These are
$B\bar{B}^*\pi$, $B^*\bar{B}^*\pi$ and $B_s^*\bar{B}_s^*$ at
$10.75\,\gev$, $10.79\,\gev$ and $10.83\,\gev$, respectively. The
$B_s^*\bar{B}_s^*$ channel is the only among the
$B_s^{(*)}\bar{B}_s^{(*)}$ channels that shows a prominent $\Uf$
signal~\cite{Abdesselam:2016tbc}. Production of the $B\bar{B}^*\pi$
and $B^*\bar{B}^*\pi$ channels proceeds entirely via intermediate
$Z_b(10610)\pi$ and $Z_b(10650)\pi$ states,
respectively~\cite{Garmash:2015rfd}, while the $Z_b\pi$ channels show
prominent $\Uf$ signals in the $\hbn\pp$ final
state~\cite{Abdesselam:2015zza}. Thus we multiply the constant width
$\Gamma$ by an energy-dependent factor
\begin{equation}
  1-x-y\ +x\,(p_1/p_1^{(0)})^3+y\,(\frac{2}{3}\,p_2/p_2^{(0)}+\frac{1}{3}\,p_3/p_3^{(0)}),
\end{equation}
where $p_i$ ($i=1,2,3$) are momenta in the $B_s^*\bar{B}_s^*$,
$Z_b(10610)\pi$ and $Z_b(10650)\pi$ pairs, respectively, and
superscript $(0)$ denotes momentum calculated for the nominal $\Uf$
mass. The factors $\frac{2}{3}$ and $\frac{1}{3}$ correspond
approximately to the ratio of the $B\bar{B}^*\pi$ and
$B^*\bar{B}^*\pi$ cross sections~\cite{Garmash:2015rfd}. We set the
weights $x$ and $y$ to various values of 0, 0.2, 0.4, and 0.6 with the
restriction $x+y\leq0.8$. We find that by introducing the effect of
the $Z_b\pi$ thresholds we always increase the significance of the new
structure. The reason for this is that the thresholds suppress the
lower mass tails of the $\Uf$ resonance, and this leads to worse
description of data under the null hypothesis.

We consider the possibility that either the $\Us$ or the new
structure, or both of them have a uniform distribution over the DP instead
of that of the $\Uf$ model. This influences the $\Gamma_f(s)$
dependence on energy and the decoherence factors. For the $\Uf-\Us$
decoherence we find 0.53, 0.66, and 0.82 in the $\Uo\pp$, $\Ut\pp$ and
$\Uth\pp$ channels, respectively, while for the $\Uf-$new structure
decoherence we find 0.44, 0.34, and 0.85.

In the default model the tails of the $\Ut\to\Uo\pp$ and
$\Uth\to\Ut\pp$ transitions are calculated under an assumption that
matrix elements of the three-body decays grow with $\mpp$ as
$\mppsq$. This growth can be damped due to the presence of some form
factor related to details of the hadronization
process~\cite{Voloshin_priv}. We consider a damping factor of the
decay amplitude in the form
\begin{equation}
  \frac{1}{1+\frac{\mpp^2}{\Lambda^2}}
  \label{eq:damp}
\end{equation}
with the parameter $\Lambda$ taking the values 1, 2 and $4\,\gev$.
The $\Gamma_f(s)$ dependence for the $\U(2S,3S)$ tails is recalculated
for each value of $\Lambda$.
In all cases we find that the significance of the new structure
increases. Thus, the default model corresponds to the fastest possible
growth of the $\U(2S,3S)$ tails with energy and this gives a
conservative estimate of the significance.
We note that another reason why the growth of the non-resonant
contribution with the energy could be damped is the crossing of
various open-flavor thresholds.
The tail of the $\Uth$ resonance can also contribute to the $\Uth\pp$
final state. Thus, we include it into the fit using the same
parametrization as for the $\Ut\pp$ channel. To calculate its energy
dependence and decoherence factors we assume that corresponding
distribution over DP is uniform.

The significance of the new structure in the $\Uth\pp$ channel in the
default model is $3.9\,\sigma$. The reason why it is so high is
evident from Fig.~\ref{fitted_xsec_3model_031118}~(lowest panel):
in the absence of the new structure the description of the $\Uth\pp$
data in the region of the $\Uf$ peak becomes very poor.  Thus, the new
structure increases flexibility of the fit function in the region of
the $\Uf$ peak due to interference. Obviously, a similar effect of
interference can be achieved with a non-resonant contribution instead
of the new structure. Indeed, the significance of the new structure in
the $\Uth\pp$ channel drops to $0.6\,\sigma$ once the non-resonant
contribution is added. The production threshold in the $\Uth\pp$
channel is above the continuum energy of $10.52\,\gev$, therefore
the information about the non-resonant contribution is limited, and
available data do not allow to study the interplay between the new
structure and the non-resonant contribution.

We consider the maximum deviation of the result as a systematic
uncertainty associated with a given source. The different
contributions and the total uncertainty obtained as their quadrature
sum are presented in Table~\ref{tab:syst_5s_6s}.
\begin{table}[htbp]
\caption{ Systematic uncertainties in the mass and width (in
  $\mevm$ and $\mev$, respectively) of the $\Uf$, $\Us$ and new
  structure. }
\renewcommand*{\arraystretch}{1.4}
\label{tab:syst_5s_6s}
\centering
\begin{tabular}{@{}lcccccc@{}} \toprule
& $\rm M_{\Uf}$ & $\Gamma_{\Uf}$ & $\rm M_{\Us}$ & $\Gamma_{\Us}$ & $\rm M_{new}$ & $\Gamma_{\rm new}$ \\ \midrule
$\U_\mathrm{new}$ parameterization & $^{+0.0}_{-0.8}$ & $^{+0.0}_{-0.4}$ & $^{+0.0}_{-0.7}$ & $^{+0.0}_{-0.6}$ & $-$ & $-$ \\
$\Uf$ parameterization & $^{+2.2}_{-0.0}$ & $^{+0.5}_{-0.0}$ & $^{+0.3}_{-0.0}$ & $\pm0.0$ & $^{+0.0}_{-0.1}$ & $^{+0.0}_{-0.1}$ \\
$\U_\mathrm{new}$, $\Us$ DP models & $^{+0.1}_{-0.4}$ & $\pm0.2$ & $^{+0.7}_{-0.8}$ & $^{+0.7}_{-0.5}$ & $^{+0.6}_{-0.0}$ & $^{+3.1}_{-0.6}$ \\
$\U(2S,3S)$ tails & $^{+0.2}_{-0.3}$ & $^{+0.0}_{-1.0}$ & $^{+0.6}_{-0.7}$ & $^{+0.0}_{-1.6}$ & $^{+0.4}_{-1.1}$ & $^{+2.3}_{-3.2}$ \\ \midrule
Total                     & $^{+2.2}_{-0.9}$ & $^{+0.5}_{-1.1}$ & $^{+1.0}_{-1.3}$ & $^{+0.7}_{-1.8}$ & $^{+0.7}_{-1.1}$ & $^{+3.9}_{-3.3}$ \\
\bottomrule
\end{tabular}
\end{table}

We find that the significance of the new structure in the $\Ut\pp$
channel and global significance combined over all channels remain
above $5.1\,\sigma$ and $5.2\,\sigma$, respectively, for all the
variations that were introduced to study systematic uncertainties.
The lowest significances are reached when we include the non-resonant
contribution into the $\Uth\pp$ channel.

\begin{figure}[htbp]
\centering
\includegraphics[width=0.65\textwidth]{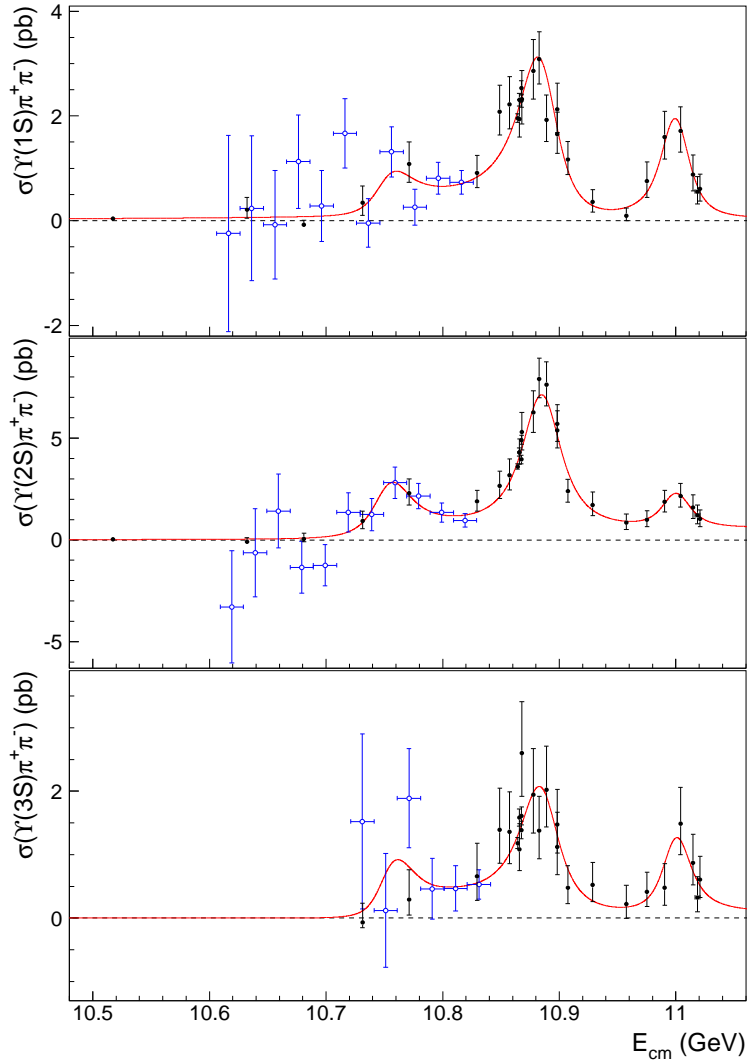}
\caption{ Energy dependence of the $\ee\to\Un\pp$ cross sections
  ($n=1,2,3$ from top to bottom). Black points with error bars and red
  solid histograms are the same as in
  Fig.~\ref{fitted_xsec_3model_031118}. Blue open dots with error bars
  are cross sections estimated using ISR process in the $\Uf$
  on-resonance data. The blue dots are intended for visualisation
  only. }
\label{fitted_xsec_w_tails_model_model_add_151218}
\end{figure}

To visualize the ISR contribution to the measurement of the cross section 
energy dependence we estimate the $\ee\to\Un\pp$ cross sections
based on the ISR tails.
For this, we divide the background-subtracted $\mmpp$ distributions in
the ISR tail regions by the ISR luminosity function and efficiency, and
apply other corrections from Eq.~(\ref{eq:sigma_born}).
Technically, to obtain the correction function we compute the $\Un\pp$
signal shape under an assumption that the cross section is constant
with the c.m.\ energy and divide the obtained shape by the $\Uf$
on-resonance cross section.
The results are presented in
Fig.~\ref{fitted_xsec_w_tails_model_model_add_151218}.
The cross sections estimated via ISR are compatible with 
the scan results and provide support for the new structure.
However, they are not intended to be used in the fit, because the
uncertainties are statistical only and do not include contributions due
to the background subtraction. Also, the ISR luminosity changes
rapidly 
in the studied energy region, which complicates accurate description
of the resolution effects. It is only in the fit to the ISR tails in
the $\mmpp$ distribution that all these issues are addressed
rigorously.

All previous measurements of the $\U(4S)\to\U(1S,2S)\pp$ branching
fractions~\cite{Guido:2017cts,Sokolov:2009zy,Aubert:2008az} assume
that the $\ee\to\U(1S,2S)\pp$ cross sections measured at the $\U(4S)$
are entirely due to $\U(4S)$ decays. To study the
implications of the presence of non-resonant contributions,
we include the $\U(4S)$ amplitude into the fit function and scan the
$-2\ln L$
in the $\U(4S)$ branching fractions.
The 67\% confidence level intervals for $\br(\U(4S)\to\Uo\pp)$ and
$\br(\U(4S)\to\Ut\pp)$ are found to be $(1.2-16)\times10^{-5}$ and
$(1.3-9.6)\times10^{-5}$, respectively.
The constraints on the branching fractions become weak because of 
interference between the $\U(4S)$ and the non-resonant amplitudes.
We also find that introducing the $\U(4S)$ into the fit has a
negligible effect on the measured mass and width values of the $\Uf$,
$\Us$, and new structure.

\section{Conclusions}
\label{sec:concl}

We report a new measurement of the energy dependence of the
$\ee\to\Un\pp$ ($n=1,2,3$) cross sections that supersedes the previous
Belle result reported in Ref.~\cite{Santel:2015qga}. We observe a new
structure in the energy dependence; if described by a Breit-Wigner
amplitude, its mass and width are found to be
$M=(10752.7\pm5.9\,^{+0.7}_{-1.1})\,\mevm$ and
$\Gamma=(35.5^{+17.6}_{-11.3}\,^{+3.9}_{-3.3})\,\mev$.
The global significance of the new structure including a systematic
uncertainty is 5.2 standard deviations.
We also report measurements of the $\Uf$ and $\Us$ parameters with 
improved accuracy.

We find evidence for the $\ee\to\Uo\pp$ process at the energy
$10.52\,\gev$.
Its cross section and the $\mpp$ distribution are consistent with the
expectations for the $\Ut$ and $\Uth$ tails.
Because of the presence of the non-resonant contributions, the
$\U(4S)\to\U(1S,2S)\pp$ branching fractions are poorly constrained by
the available cross section measurements that are all performed at the
$\U(4S)$ peak.

The new structure could have a resonant origin and correspond to a
signal for the not yet observed $\U(3D)$ state provided $S-D$ mixing is
enhanced~\cite{Badalian:2009bu}, or an exotic state, e.g. a compact
tetraquark~\cite{Ali:2009es} or
hadrobottomonium~\cite{Dubynskiy:2008mq}. It could also be a
non-resonant effect due to some complicated rescattering.
Information on the cross section energy dependence for more channels, 
with both hidden and open $b$ flavor, is needed to clarify the nature
of the new structure.

\acknowledgments

We are grateful to A.G.~Shamov for preparing for us the generator for
QED production of four-track final states~\cite{Krachkov:2019kty} and
to M.B.~Voloshin for fruitful discussions.
We thank the KEKB group for the excellent operation of the
accelerator; the KEK cryogenics group for the efficient
operation of the solenoid; and the KEK computer group, and the Pacific Northwest National
Laboratory (PNNL) Environmental Molecular Sciences Laboratory (EMSL)
computing group for strong computing support; and the National
Institute of Informatics, and Science Information NETwork 5 (SINET5) for
valuable network support.  We acknowledge support from
the Ministry of Education, Culture, Sports, Science, and
Technology (MEXT) of Japan, the Japan Society for the 
Promotion of Science (JSPS), and the Tau-Lepton Physics 
Research Center of Nagoya University; 
the Australian Research Council including grants
DP180102629, 
DP170102389, 
DP170102204, 
DP150103061, 
FT130100303; 
Austrian Science Fund (FWF);
the National Natural Science Foundation of China under Contracts
No.~11435013,  
No.~11475187,  
No.~11521505,  
No.~11575017,  
No.~11675166,  
No.~11705209;  
Key Research Program of Frontier Sciences, Chinese Academy of Sciences (CAS), Grant No.~QYZDJ-SSW-SLH011; 
the  CAS Center for Excellence in Particle Physics (CCEPP); 
the Shanghai Pujiang Program under Grant No.~18PJ1401000;  
the Ministry of Education, Youth and Sports of the Czech
Republic under Contract No.~LTT17020;
the Carl Zeiss Foundation, the Deutsche Forschungsgemeinschaft, the
Excellence Cluster Universe, and the VolkswagenStiftung;
the Department of Science and Technology of India; 
the Istituto Nazionale di Fisica Nucleare of Italy; 
National Research Foundation (NRF) of Korea Grants
No.~2015H1A2A1033649, No.~2016R1D1A1B01010135, No.~2016K1A3A7A09005
603, No.~2016R1D1A1B02012900, No.~2018R1A2B3003 643,
No.~2018R1A6A1A06024970, No.~2018R1D1 A1B07047294; Radiation Science Research Institute, Foreign Large-size Research Facility Application Supporting project, the Global Science Experimental Data Hub Center of the Korea Institute of Science and Technology Information and KREONET/GLORIAD;
the Polish Ministry of Science and Higher Education and 
the National Science Center;
the Russian Science Foundation (RSF), Grant No. 18-12-00226;
the Slovenian Research Agency;
Ikerbasque, Basque Foundation for Science, Spain;
the Swiss National Science Foundation; 
the Ministry of Education and the Ministry of Science and Technology of Taiwan;
and the United States Department of Energy and the National Science Foundation.

\end{document}

%% file: pub543.tex
\affiliation[1]{University of the Basque Country UPV/EHU, 48080 Bilbao}
\affiliation[2]{Brookhaven National Laboratory, Upton, New York 11973}
\affiliation[3]{Budker Institute of Nuclear Physics SB RAS, Novosibirsk 630090}
\affiliation[4]{Faculty of Mathematics and Physics, Charles University, 121 16 Prague}
\affiliation[5]{Chonnam National University, Kwangju 660-701}
\affiliation[6]{University of Cincinnati, Cincinnati, Ohio 45221}
\affiliation[7]{Deutsches Elektronen--Synchrotron, 22607 Hamburg}
\affiliation[8]{Key Laboratory of Nuclear Physics and Ion-beam Application (MOE) and Institute of Modern Physics, Fudan University, Shanghai 200443}
\affiliation[9]{Justus-Liebig-Universit\"at Gie\ss{}en, 35392 Gie\ss{}en}
\affiliation[10]{Gifu University, Gifu 501-1193}
\affiliation[11]{II. Physikalisches Institut, Georg-August-Universit\"at G\"ottingen, 37073 G\"ottingen}
\affiliation[12]{SOKENDAI (The Graduate University for Advanced Studies), Hayama 240-0193}
\affiliation[13]{Gyeongsang National University, Chinju 660-701}
\affiliation[14]{Hanyang University, Seoul 133-791}
\affiliation[15]{University of Hawaii, Honolulu, Hawaii 96822}
\affiliation[16]{High Energy Accelerator Research Organization (KEK), Tsukuba 305-0801}
\affiliation[17]{J-PARC Branch, KEK Theory Center, High Energy Accelerator Research Organization (KEK), Tsukuba 305-0801}
\affiliation[18]{Forschungszentrum J\"{u}lich, 52425 J\"{u}lich}
\affiliation[19]{IKERBASQUE, Basque Foundation for Science, 48013 Bilbao}
\affiliation[20]{Indian Institute of Science Education and Research Mohali, SAS Nagar, 140306}
\affiliation[21]{Indian Institute of Technology Bhubaneswar, Satya Nagar 751007}
\affiliation[22]{Indian Institute of Technology Guwahati, Assam 781039}
\affiliation[23]{Indian Institute of Technology Hyderabad, Telangana 502285}
\affiliation[24]{Indian Institute of Technology Madras, Chennai 600036}
\affiliation[25]{Indiana University, Bloomington, Indiana 47408}
\affiliation[26]{Institute of High Energy Physics, Chinese Academy of Sciences, Beijing 100049}
\affiliation[27]{Institute of High Energy Physics, Vienna 1050}
\affiliation[28]{Institute for High Energy Physics, Protvino 142281}
\affiliation[29]{INFN - Sezione di Napoli, 80126 Napoli}
\affiliation[30]{INFN - Sezione di Torino, 10125 Torino}
\affiliation[31]{Advanced Science Research Center, Japan Atomic Energy Agency, Naka 319-1195}
\affiliation[32]{J. Stefan Institute, 1000 Ljubljana}
\affiliation[33]{Institut f\"ur Experimentelle Teilchenphysik, Karlsruher Institut f\"ur Technologie, 76131 Karlsruhe}
\affiliation[34]{Kennesaw State University, Kennesaw, Georgia 30144}
\affiliation[35]{King Abdulaziz City for Science and Technology, Riyadh 11442}
\affiliation[36]{Kitasato University, Sagamihara 252-0373}
\affiliation[37]{Korea Institute of Science and Technology Information, Daejeon 305-806}
\affiliation[38]{Korea University, Seoul 136-713}
\affiliation[39]{Kyoto University, Kyoto 606-8502}
\affiliation[40]{Kyungpook National University, Daegu 702-701}
\affiliation[41]{LAL, Univ. Paris-Sud, CNRS/IN2P3, Universit\'{e} Paris-Saclay, Orsay}
\affiliation[42]{\'Ecole Polytechnique F\'ed\'erale de Lausanne (EPFL), Lausanne 1015}
\affiliation[43]{P.N. Lebedev Physical Institute of the Russian Academy of Sciences, Moscow 119991}
\affiliation[44]{Faculty of Mathematics and Physics, University of Ljubljana, 1000 Ljubljana}
\affiliation[45]{Ludwig Maximilians University, 80539 Munich}
\affiliation[46]{Luther College, Decorah, Iowa 52101}
\affiliation[47]{University of Malaya, 50603 Kuala Lumpur}
\affiliation[48]{University of Maribor, 2000 Maribor}
\affiliation[49]{Max-Planck-Institut f\"ur Physik, 80805 M\"unchen}
\affiliation[50]{School of Physics, University of Melbourne, Victoria 3010}
\affiliation[51]{University of Mississippi, University, Mississippi 38677}
\affiliation[52]{University of Miyazaki, Miyazaki 889-2192}
\affiliation[53]{Moscow Physical Engineering Institute, Moscow 115409}
\affiliation[54]{Moscow Institute of Physics and Technology, Moscow Region 141700}
\affiliation[55]{Graduate School of Science, Nagoya University, Nagoya 464-8602}
\affiliation[56]{Kobayashi-Maskawa Institute, Nagoya University, Nagoya 464-8602}
\affiliation[57]{Universit\`{a} di Napoli Federico II, 80055 Napoli}
\affiliation[58]{Nara Women's University, Nara 630-8506}
\affiliation[59]{National Central University, Chung-li 32054}
\affiliation[60]{National United University, Miao Li 36003}
\affiliation[61]{Department of Physics, National Taiwan University, Taipei 10617}
\affiliation[62]{H. Niewodniczanski Institute of Nuclear Physics, Krakow 31-342}
\affiliation[63]{Nippon Dental University, Niigata 951-8580}
\affiliation[64]{Niigata University, Niigata 950-2181}
\affiliation[65]{Novosibirsk State University, Novosibirsk 630090}
\affiliation[66]{Osaka City University, Osaka 558-8585}
\affiliation[67]{Pacific Northwest National Laboratory, Richland, Washington 99352}
\affiliation[68]{University of Pittsburgh, Pittsburgh, Pennsylvania 15260}
\affiliation[69]{Punjab Agricultural University, Ludhiana 141004}
\affiliation[70]{Research Center for Nuclear Physics, Osaka University, Osaka 567-0047}
\affiliation[71]{Theoretical Research Division, Nishina Center, RIKEN, Saitama 351-0198}
\affiliation[72]{University of Science and Technology of China, Hefei 230026}
\affiliation[73]{Seoul National University, Seoul 151-742}
\affiliation[74]{Showa Pharmaceutical University, Tokyo 194-8543}
\affiliation[75]{Soongsil University, Seoul 156-743}
\affiliation[76]{Stefan Meyer Institute for Subatomic Physics, Vienna 1090}
\affiliation[77]{Sungkyunkwan University, Suwon 440-746}
\affiliation[78]{School of Physics, University of Sydney, New South Wales 2006}
\affiliation[79]{Department of Physics, Faculty of Science, University of Tabuk, Tabuk 71451}
\affiliation[80]{Tata Institute of Fundamental Research, Mumbai 400005}
\affiliation[81]{Department of Physics, Technische Universit\"at M\"unchen, 85748 Garching}
\affiliation[82]{Toho University, Funabashi 274-8510}
\affiliation[83]{Department of Physics, Tohoku University, Sendai 980-8578}
\affiliation[84]{Earthquake Research Institute, University of Tokyo, Tokyo 113-0032}
\affiliation[85]{Department of Physics, University of Tokyo, Tokyo 113-0033}
\affiliation[86]{Tokyo Metropolitan University, Tokyo 192-0397}
\affiliation[87]{Virginia Polytechnic Institute and State University, Blacksburg, Virginia 24061}
\affiliation[88]{Wayne State University, Detroit, Michigan 48202}
\affiliation[89]{Yamagata University, Yamagata 990-8560}
\affiliation[90]{Yonsei University, Seoul 120-749}
  \author[43,54]{R.~Mizuk,} 
  \author[3,65]{A.~Bondar,} 
  \author[16,12]{I.~Adachi,} 
  \author[85]{H.~Aihara,} 
  \author[2]{D.~M.~Asner,} 
  \author[3,65]{V.~Aulchenko,} 
  \author[54]{T.~Aushev,} 
  \author[79]{R.~Ayad,} 
  \author[79,35]{I.~Badhrees,} 
  \author[21]{S.~Bahinipati,} 
% ...
\collaboration{The Belle Collaboration}